\documentclass[1p]{elsarticle}
\usepackage[english]{babel}
\usepackage{amsmath,amsthm}
\usepackage{multirow}
\usepackage{bbold}
\usepackage{tikz}
\usepgflibrary{arrows.meta}
\usepackage{latexsym}
\usepackage{color}
\usepackage{graphicx}

\catcode `\@=11 \@addtoreset{equation}{section}
\def\theequation{\arabic{section}.\arabic{equation}}
\catcode `\@=12

\newcommand{\nn}{\nonumber}

\newcommand{\dd}{\mathrm{d}} 

\newcommand{\M}[1]{\mathbf{#1}}

\DeclareMathOperator{\tr}{\mathrm{tr}}

\newcommand{\RR}{\mathbb{R}}
\newcommand{\CC}{\mathbb{C}}

\newcommand{\ee}{\mathrm{e}}
\newcommand{\ii}{\mathrm i}

\DeclareMathOperator{\sign}{\mathrm{sign}}

\DeclareMathOperator{\IM}{\mathrm{Im}}

\newcommand{\PHI}{\M{\phi}}

\renewcommand{\thesection}{\Roman{section}}

\newcommand{\be}{\begin{equation}}
\newcommand{\en}{\end{equation}}
\newcommand{\bea}{\begin{eqnarray}}
\newcommand{\ena}{\end{eqnarray}}
\newcommand{\beano}{\begin{eqnarray*}}
\newcommand{\enano}{\end{eqnarray*}}
\newcommand{\bee}{\begin{enumerate}}
\newcommand{\ene}{\end{enumerate}}

\newcommand{\1}{1 \!\! 1}

\begin{document}

\title{Pseudo-Hermitian Random Matrix Models: General Formalism}


\author[1]{Joshua Feinberg\fnref{orcid}}
\author[1]{Roman Riser}
\address[1]{Department of Physics\\and\\Haifa Research Center for Theoretical Physics and Astrophysics\\University of Haifa, Haifa 31905, Israel}
\fntext[orcid]{https://orcid.org/0000-0002-2869-0010}



\begin{abstract}
Pseudo-hermitian matrices are matrices hermitian with respect to an indefinite metric. They can be thought of as the truncation of pseudo-hermitian operators, defined over some Krein space, together with the associated metric, to a finite dimensional subspace. As such, they can be used, in the usual spirit of random matrix theory, to model chaotic or disordered $PT$-symmetric quantum systems, or their gain-loss-balanced classical analogs, in the phase of broken $PT$-symmetry. The eigenvalues of pseudo-hermitian matrices are either real, or come in complex-conjugate pairs. In this paper we introduce a family of pseudo-hermitian random matrix models, depending parametrically on their metric. We apply the diagrammatic method to obtain its averaged resolvent and density of eigenvalues as explicit functions of the metric, in the limit of large matrix size $N$. As a concrete example, which is essentially an ensemble of elements of the non-compact unitary Lie algebra, we choose a particularly simple set of metrics, and compute the resulting resolvent and density of eigenvalues in closed form. The spectrum consists of a finite fraction of complex eigenvalues, which occupy uniformly two two-dimensional blobs, symmetric with respect to the real axis, as well as the complimentary fraction of real eigenvalues, condensed in a finite segment, with a known non-uniform density. The numbers  of complex and real eigenvalues depend on the signature of the metric, that is, the numbers of its positive and negative eigenvalues. We have also carried thorough numerical analysis of the model for these particular metrics. Our numerical results converge rapidly towards the asymptotic analytical large-$N$ expressions. \end{abstract}
\vspace{1cm}

\maketitle

\newpage

\section{Introduction}

PT-symmetric quantum mechanics (PTQM) \cite{BB} and its broader applications (see \cite{CMB} for recent reviews) have been at the focus of intensive and prolific research activity during the past twenty-five years or so. Generally speaking, the Hilbert space of a PT-symmetric quantum mechanical model is endowed with a non-trivial metric operator, with respect to which the hamiltonian is hermitian. Hamiltonians of PTQM models with proper positive metrics are sometimes referred to as {\it quasi-hermitian} \cite{Dieudonne, stellenbosch}. They are diagonalizable, and their spectrum is essentially real, because they are similar to a conventionally hermitian hamiltonian. In contrast, hamiltonians of PTQM models with {\it indefinite} metrics are referred to as {\it pseudo-hermitian} \cite{Froissart, Mostafazadeh}, and their eigenvalues are either real or come in  complex-conjugate pairs. Pseudo-hermitian hamiltonians describe systems with {\em broken} $PT-$symmetry \cite{CMB}. For a mathematically precise summary of the nomenclature of quasi-hermiticity and pseudo-hermiticity see \cite{Fring-Assis}.

{\em Quasi-hermitian} (QH) matrices can be thought of as truncated quasi-hermitian linear operators into finite dimensional space. More precisely, we say that the $N\times N$ matrix $\M{\phi}$ (the ``hamiltonian") is quasi-hermitian with respect to the $N\times N$ hermitian proper (non-negative) metric $\M{B}$, if $\M{\phi}$ and its adjoint fulfill the {\em intertwining relation}
\begin{equation}\label{intertwining}
\M{\phi}^\dagger \M{B} = \M{B}\M{\phi} \,.
\end{equation} 
The relation \eqref{intertwining} simply means that $\M{\phi}$ is hermitian in a vector space endowed with a non-trivial metric $\M{B}$. (For $\M{B}=\mathbb{1}$ it reduces of course to ordinary hermiticity.)

Following \cite{FRDecember, MK, FR-review}, we should make further distinction between QH matrices and {\it strictly}-quasi-hermitian (SQH) matrices.  SQH matrices are hermitian with respect to positive definite (and therefore invertible) metrics $\M{B}$, in contrast to merely QH matrices, whose associated non-negative metrics may be non-invertible. 

SQH random matrices arise, for example, in studying the spectrum of stable small oscillations of large mechanical systems with large connectivity \cite{FRDecember, MK}. The low-frequency part of the spectrum of such systems was found in \cite{FRDecember, MK} to be universal, with the phonon density of states tending to a non-zero constant at zero frequency. In other words, phonons in such systems have universal spectral dimension $d_s = 1$.

An interesting SQH random matrix model was introduced in \cite{JK}. These authors fixed a metric $\M{B}$, and took the hamiltonian $\M{\phi}$ as random, with the aim of studying numerically the dependence of the average density of eigenvalues and level spacing statistics on the metric. Yet another interesting example of an SQH random matrix model, akin to the Dicke model of superradiance, was provided by \cite{DGK}, in which a numerical study of the level spacing distribution was carried out.

The difference of the two sides of \eqref{intertwining} is an anti-hermitian matrix. Thus, given a generic {\em invertible} hermitian metric $\M{B}$, \eqref{intertwining} amounts to a system of $N^2$ independent real equations for the $2N^2$ real parameters of $\PHI$, resulting in $N^2$ linearly independent solutions. Indeed, as was pointed out in \cite{carlson} (and in the physics literature in \cite{JK}), the general solution of \eqref{intertwining} for $\PHI$ is 
\begin{equation} \label{eq:A}
\PHI=\M{A}\M{B},
\end{equation}
where $\M{A}=\M{A}^\dagger$ is a hermitian matrix.  Therefore, given $\M{B}$, there are $N^2$ linearly independent matrices which are SQH with respect to $\M{B}$. 

An important corollary of \eqref{eq:A} is that for a positive definite metric, $\PHI$ is similar to a hermitian matrix, with the similarity matrix being essentially the (invertible) square root of $\M{B}$. (The latter is well defined and hermitian, because $\M{B}$ is positive definite.) Consequently, SQH matrices are diagonalizable and have purely real eigenvalue spectrum. This can be seen easily when we write 
\begin{equation}
\PHI=\sqrt{\M{B}}^{\ -1}\underbrace{\sqrt{\M{B}}\M{A}\sqrt{\M{B}}}_{\text{hermitian}} \sqrt{\M{B}}.
\end{equation}

Upon truncation to finite vector spaces, pseudo-hermitian operators beget {\em pseudo-hermitian} (PH) matrices. Such matrices are still defined by the intertwining relation \eqref{intertwining}, but now the metric $\M{B}$ has indefinite signature.

The dimensionality of the space of solutions of \eqref{intertwining} for $\PHI$ is clearly independent of the signature of the metric $\M{B}$, as long as it is invertible. Thus, since the number of independent solutions of the form \eqref{eq:A} saturates this dimensionality, it is the general solution of \eqref{intertwining} {\em regardless} of the signature of $\M{B}$. Consequently, any matrix $\PHI$ which is PH with respect to an indefinite invertible metric $\M{B}$ can be written in the form \eqref{eq:A}, with $\M{A}$ being a hermitian matrix. 

A PH matrix, in contrast to an SQH matrix, need not\footnote{A PH matrix $\M{\phi}$ may still have purely real eigenvalues. For example, $\M{\phi}=1\!\!1$ is PH with respect to a metric $\M{B}$ of any signature. A real diagonal $\M{\phi}$ and a diagonal $\M{B}$ of any signature is yet another example.} be similar to a hermitian matrix. In this case, since according to \eqref{intertwining}  $\M{\phi}$ is similar to its hermitian adjoint, it follows that the characteristic polynomial of $\M{\phi}$ has real coefficients: $\left(\det (z-\M{\phi})\right)^* = \det (z^*-\M{\phi}).$ Thus, the eigenvalues of $\M{\phi}$ are either real, or come in complex-conjugate pairs. See \cite{KA} for a recent discussion of (real asymmetric) PH random matrices.

For the sake of completeness, let us briefly discuss the case of a singular metric $\M{B}$. Thus, assume ${\rm rank}(\M{B}) = p<N$. Then, by transforming to a basis in which $\M{B}$ is diagonal, it is easy to see that \eqref{intertwining} amounts to a system of $N^2-(N-p)^2$ independent real equations  for the $2N^2$ real parameters of $\PHI$, resulting in $N^2 + (N-p)^2$ linearly independent solutions. In contrast, it is easy to check that solutions of the form \eqref{eq:A} comprise in this case only a subspace of real dimension $N^2-(N-p)^2$. The complementary subspace of solutions, which cannot be expressed in the form \eqref{eq:A}, has therefore real dimension $2(N-p)^2$, namely, the dimension of an $(N-p)\times (N-p)$ complex matrix.  This means that not all complex eigenvalues of $\PHI$ in this case would come in complex conjugate pairs. We shall not pursue here the study of such matrices any further, since in this paper we focus exclusively on invertible metrics $\M{B}$. 

We should also mention in passing that some special PH matrices may be non-diagonalizable, as the reader can easily check by working out all the matrices which are PH with respect to the metric $\M{B} = \sigma_3 = {\rm diag}(1,-1)$. In this case, matrices with degenerate eigenvalues and whose diagonal elements are not equal, admit only a Jordan form and are not diagonalizable. Such non-diagonalizable matrices clearly form a set of zero measure in the space of all PH matrices with respect to a given indefinite metric, and need not concern us in our statistical analysis of the random matrix models we study in this paper.

In this paper we introduce and study a family of pseudo-hermitian random matrix models, which depend parametrically on a fixed (deterministic), invertible, indefinite metric $\M{B}$.  
The hermitian matrix $\M{A} = \PHI\M{B}^{-1}$ in \eqref{eq:A} is uniquely determined, and is the source of randomness. Thus, the probability ensemble of the PH matrices $\PHI$ is induced from that of the hermitian matrices $\M{A}$. We are free to choose the latter at will, and a natural choice is to draw $\M{A}$ from the Gaussian Unitary Ensemble (GUE)
\begin{equation}\label{eq:GUE}
P(\M{A})=\frac{1}{Z_N} \ee^{-\frac{N m^2}{2} \tr \M{A}^2},
\end{equation}
where $m>0$ is a parameter and $Z_N$ is the normalization constant. Thus, \eqref{eq:GUE} induces a PDF on the $2N^2$ real parameters of $\PHI$,
\begin{equation}\label{ensemble}
P(\PHI)=\frac{1}{\tilde{Z}_N} \ee^{-\frac{N m^2}{2} \tr (\M{B}^{-2} \PHI^\dagger \PHI) }\delta\left(\PHI^\dagger\M{B} - \M{B}\PHI\right),
\end{equation}
where $\tilde{Z}_N$ is another normalization constant. Therefore, an $N\times N$ complex matrix $\PHI$ is drawn with the Gaussian weight in \eqref{ensemble}, and then the remaining factor $\delta\left(\PHI^\dagger\M{B} - \M{B}\PHI\right)$, which is supported over the $N^2$ independent real and imaginary parts of the anti-hermitian matrix $\PHI^\dagger\M{B} - \M{B}\PHI$, filters in those matrices which are PH with respect to the metric $\M{B}$. This is the matrix model we focus on, and our main goal in this paper is to determine explicitly the ensemble-averaged density of the  eigenvalues of $\PHI$ on the real axis and in the complex eigenvalues in the limit $N\rightarrow\infty$.

We have also studied numerically the spectral statistics of $\PHI$ for ensembles of $\M{A}$ other than  \eqref{eq:GUE}. Our results \cite{FRuniversality} indicate that many attributes of the spectral statistics of $\PHI$ are universal, and depend, perhaps not surprisingly, only on the variance of fluctuations of $\M{A}$.

We have recently introduced and previewed the model \eqref{ensemble} in \cite{FR-review}, and the purpose of the present paper is to provide detailed exposition of our analysis. We derive explicit equations for determining the averaged resolvent and density of eigenvalues as explicit functions of the metric, in the limit of large matrix size $N$, by means of the diagrammatic method. As a concrete example, we set the metric to be diagonal with entries $+1$ or $-1$, and compute the resulting resolvent and density of eigenvalues in closed form. The spectrum in this case  consists of a finite fraction of complex eigenvalues, which occupy uniformly two two-dimensional blobs of known shape, symmetric with respect to the real axis, as well as the complimentary fraction of real eigenvalues, condensed in a finite segment, with a known non-uniform density. The numbers  of complex and real eigenvalues depend on the signature of the metric, that is, the numbers of its positive and negative eigenvalues, both of which are assumed to be finite fractions of $N$. 

The average spectrum thus obtained for a particular such metric, is essentially the spectrum of a randomly chosen generator of the non-compact unitary group with the same signature. Rotating this spectrum by ninety degrees in the complex plane gives us the spectrum of a randomly chosen generator of the corresponding non-compact orthogonal group.  

We have also carried thorough numerical analysis of the model for these particular metrics. Our numerical results converge rapidly towards the asymptotic analytical large-$N$ expressions. 

This is perhaps a good place to mention that products of the form \eqref{eq:A}, with $\M{A}$ drawn from the GUE and a deterministic hermitian matrix $\M{B}$,  appear also in random matrix theory with an external source \cite{BH, BH1}, in studying the eigenvalue  statistics\footnote{The combination $\PHI=\M{A}\M{B}$ appears in this context because of the shift ${\rm tr}\M{A}^2\mapsto {\rm tr} ( \M{A} + \M{B})^2$ in the exponential of the Gaussian weight, as can be seen e.g. in Eq.(3.1) in \cite{BH}.} of the {\em hermitian}   matrix $\M{H} = \M{A} + \M{B}$. This is in contrast to the spectral statistics of the {\em non-hermitian} (or more precisely, PH) random matrix $\PHI=\M{A}\M{B}$ studied in the present paper.

The rest of the paper is organized as follows: In Section \ref{sec:methods} we apply the method of hermitization and the diagrammatic method to deriving the self-consistent gap equations in the large-$N$, planar limit. We then analyze the gap equations in Section \ref{gap-dos} and determine from them the phase structure of our model in the complex plane, which consists of the so-called non-holomorphic phase in a two-dimensional region ${\mathcal D}\subset\CC$ in which the eigenvalues of our PH matrix $\PHI$ condense in the large-$N$ limit, and a holomorphic phase, in the complementary domain of the complex plane. In particular, we derive explicit expressions for the Green's function of $\PHI$ in the holomorphic phase (see Eqs. \eqref{Ghol} and \eqref{Ghol1}), and in the non-holomorphic phase (see Eq. \eqref{Green-nh-final}). The discontinuity of this function in the holomorphic phase across the real axis determines the density of real eigenvalues of $\PHI$, while the density of complex eigenvalues of $\PHI$ can be determined from that Green's function in the non-holomorphic phase by applying Gauss' law, as explained in \eqref{rhoC}. Section \ref{gap-dos} is very detailed and contains many analytical results and interrelations between various quantities for generic metrics $\M{B}$. (See, e.g., \eqref{invariant} and \eqref{invariant-nh}.) Finally, in Section \ref{sec:model B} we apply our general formalism to the concrete case of the metric $\M{B}=\mathrm{diag}(1,\ldots,1,-1,\ldots,-1)$ with $k$ ones and $N-k$ minus ones, as described above, and provide explicit analytical and numerical results for the density of eigenvalues on the real axis and in the complex plane (see Eqs.~\eqref{eq:rhoreal} and \eqref{eq:rpm}, \eqref{eq:rhoC} respectively). We then discuss the relevance of these results to the non-compact Lie algebras $su(k,N-k)$ and $so(k,N-k)$. Many technical and mathematical details are relegated to the four appendices at the end of this paper.

\section{Hermitization and Large-$N$ Analysis}\label{sec:methods}
In practice, it is more convenient to work with the representation $\PHI=\M{A}\M{B}$  of the pseudo-hermitian matrix in \eqref{eq:A}, rather than with the singular PDF \eqref{ensemble}. The averaged density of eigenvalues of $\PHI$ in the complex-$w$ plane can be calculated from the Green's function (resolvent) 
\begin{equation}\label{eq:Green}
G(w)=\left\langle \frac{1}{N} \tr {1\over w -\M{A}\M{B}} \right\rangle
\end{equation}
as we explain in Section \ref{gap-dos}. Here, and from here on, angular brackets denote averaging of $\M{A}$ over the GUE ensemble \eqref{eq:GUE}. 

Averaging over $\M{A}$ becomes simpler if one can avoid the product of matrices $\M{A}\M{B}$ by the method introduced in \cite{BJN}, following which we introduce the $2N \times 2N$ block matrix 
\begin{equation}\label{H2N}
\M{H}=\left(\begin{array}{cc} 
0&\M{A}\\\M{B}&0 
\end{array}\right),
\end{equation}
which puts $\M{A}$ and $\M{B}$ in two different blocks. The resolvent of \eqref{H2N} is readily computed as 
\begin{equation}\label{resolvent}
\frac{1}{z-\M{H}}=\left(\begin{array}{cc} 
\frac{z}{z^2-\M{A}\M{B}}& \M{A}\frac{1}{z^2-\M{B}\M{A}}\\
\M{B}\frac{1}{z^2-\M{A}\M{B}}& \frac{z}{z^2-\M{B}\M{A}}
\end{array}\right),
\end{equation}
in which the upper and lower diagonal blocks are essentially the resolvents of  $\PHI = \M{AB}$ and $\PHI^\dagger = \M{BA}$, evaluated at $w=z^2$.
Thus, we arrive at the Green's function 
\begin{equation}\label{eq:Gtilde}
\widetilde{G}(z)= \left\langle \frac{1}{2N} \tr \frac{1}{z-\M{H}} \right\rangle = z G(z^2)\,,
\end{equation}
where in the last step we have used equality of the traces of the two diagonal blocks of \eqref{resolvent} due to isospectrality of $\PHI = \M{AB}$ and $\PHI^\dagger = \M{BA}$. Therefore, we can deduce the desired Green's function \eqref{eq:Green} from \eqref{eq:Gtilde}, which is more amenable to diagrammatic expansion, in a straightforward manner. 

Since $\M{H}$ is typically non-hermitian, it might have complex eigenvalues. In Appendix \ref{symmetries} we discuss the symmetries the spectrum of $\M{H}$ in the complex plane. In the large-$N$ limit, these eigenvalues  can be dense in a two-dimensional domain $\widetilde{\cal D}\subset\CC$, rendering $\widetilde{G}$ not an analytic function\footnote{More precisely, this is true only if $\M{B}$ has indefinite signature. In contrast, for positive-definite $\M{B}$, $\PHI = \M{AB}$ is SQH and therefore similar to a hermitian matrix, rendering $\widetilde{G}(z)$ possibly non-analytic only on the real axis in the complex $w=z^2$ plane.} of $z$, depending both on $z$ and $z^*$. As such, $\widetilde{G}$, or for that matter, any of the averaged blocks of \eqref{resolvent}, cannot be determined everywhere in the complex plane $\CC$ from the moments of $\M{H}$, which amounts to expanding them in inverse powers of $z$. The latter so-called perturbative Born series would converge only in the analyticity domains of  \eqref{eq:Gtilde}, which excludes $\widetilde{\cal D}$. This is in contrast with resolvents of hermitian random matrices, which are always analytic functions of $z$ everywhere off the real axis, where they can be determined, in principle, from their perturbative Born series. 

As is well known, this difficulty in computing perturbatively averaged resolvents of non-hermitian matrices by resumming their Born series, is overcome by employing the \emph{Method of Hermitization} \cite{FZ,JNGZ,CW,E97}. According to this method we can reduce the difficult problem of averaging \eqref{resolvent} over the random matrix $\M{A}$ to the more familiar problem of averaging the resolvent $\hat{\mathcal{G}}$ of the $4N\times 4N$ \emph{hermitian} matrix 
\begin{equation}\label{eq:herm}
\hat{\mathcal{H}}  = \left(\begin{array}{cc} 
0&z-\M{H}\\z^*-\M{H}^\dagger&0 
\end{array}\right)\,,
\end{equation}
namely\footnote{The block structure of \eqref{eq:G1} is isomorphic, of course, to that of  \eqref{resolvent}, with $-(z-\M{H})$ and $-(z-\M{H})^\dagger$, respectively, playing the roles of $\M{A}$ and $\M{B}$.}, 
\begin{equation}\label{eq:G1} 
\hat{\mathcal{G}}(\eta; z, z^*)= {1\over \eta - \hat{\mathcal{H}}} = 
\left(\begin{array}{cc} {\eta \over \eta^2 - (z-\M{H})(z-\M{H})^\dagger} & -(z-\M{H}) {1\over \eta^2 - (z-\M{H})^\dagger (z-\M{H}) }\\ -(z - \M{H})^\dagger {1\over \eta^2 - (z-\M{H})(z-\M{H})^\dagger} & {\eta \over \eta^2 - (z-\M{H})^\dagger (z-\M{H})}\end{array}\right).
\end{equation}

Therefore, off the real axis in the complex $\eta$-plane,  the averaged resolvent $\langle\hat{\mathcal{G}}\rangle$ is an analytic function of the spectral parameter $\eta$. This will allow us to compute $\langle\hat{\mathcal{G}}\rangle$ by expanding it diagrammatically in powers $1/\eta$, and then resum the series. This series would then converge to $\langle\hat{\mathcal{G}}\rangle$ everywhere off the real axis in the complex-$\eta$ plane.

After resumming the series, we could let $\eta\rightarrow 0$. The lower left block of \eqref{eq:G1} would then clearly converge to the desired average $\langle{1\over z-\M{H}}\rangle$ of \eqref{resolvent}, while the upper right block of \eqref{eq:G1} would converge to its adjoint. In contrast, the diagonal blocks would suffer a jump discontinuity in the limit, depending on whether one approaches the real $\eta$ axis from above or from below. More specifically, if we set  $\eta = is, s\in\RR$, then as 
$s\rightarrow 0$, 
\begin{eqnarray}\label{etas}
\lim_{s\to 0\pm} \langle\hat{\mathcal{G}}(is; z, z^*)\rangle =\nonumber\\{}\nonumber\\
&&\!\!\!\!\!\!\!\!\!\!\!\!\!\!\!\!\!\!\!\!\!\!\!\!\!\!\!\!\!\!\!\!\!\!\!\!\!\!\!\!\!\!\!\!\!\!\!\!\!\!\left(\begin{array}{cc} \mp i\pi \left\langle\delta\left(\sqrt{(z-\M{H})(z-\M{H})^\dagger}\right)\right\rangle  &  \left\langle{1 \over  z^*-\M{H}^\dagger}\right\rangle \\ {} & {} \\ \left\langle{1\over   z-\M{H} }\right\rangle & \mp i\pi\left\langle\delta\left(\sqrt{(z-\M{H})^\dagger (z-\M{H})}\right)\right\rangle \end{array}\right).
\end{eqnarray}
Thus, the diagonal blocks of \eqref{etas} converge in the limit to two isospectral anti-hermitian matrices, while the limiting off-diagonal blocks are hermitian conjugate of each other. More detailed analysis of these diagonal blocks (before averaging over $\M{A}$) is given in Appendix \ref{sec:diagonals}.

In our subsequent analysis we will have to refer to  the individual $N\times N$ blocks of  
\begin{equation}\label{eq:G} 
\hat{\mathcal{G}}(\eta; z, z^*)=\left[\left(\begin{array}{cc}
\eta&z\\z^*&\eta 
\end{array}\right)-\left(\begin{array}{cc} 
0&\M{H}\\\M{H}^\dagger&0 
\end{array}\right)\right]^{-1}=\left(\begin{array}{cccc}
\eta&0&z&-\M{A}\\
0&\eta&-\M{B}&z\\
z^*&-\M{B}&\eta&0\\
-\M{A}&z^*&0&\eta
\end{array}\right)^{-1}.
\end{equation}
Let us therefore denote these blocks as 
\begin{equation}\label{blocks}
\hat{\mathcal{G}}_{\alpha\beta}\,,\quad \alpha, \beta :=1,2,3,4     
\end{equation}
Matrix elements within a given block will be indexed by Latin indices as 
\begin{equation}\label{Latin}
\left(\hat{\mathcal{G}}_{\alpha\beta}\right)_{ij},\quad i,j  :=1,\ldots , N   
\end{equation}

Following 't Hooft, we will use Feynman diagrams with double-lined matrix (`gluon') propagators \cite{thooft} in the large-$N$ planar limit \cite{QFTNut}. To this end we expand the resolvent $\langle\hat{\mathcal{G}}(\eta;z,z^*)\rangle$ in powers of the bare `quark' propagator
\begin{equation}\label{bareprop}
\hat{\mathcal{G}}_0=\hat{\mathcal{G}}_{|_{A=0}}= \left(\begin{array}{cccc}
\eta&0&z& 0\\
0&\eta&-\M{B}&z\\
z^*&-\M{B}&\eta&0\\
 0&z^*&0&\eta
\end{array}\right)^{-1}= \begin{tikzpicture}[scale=0.3,line width=0.4pt,baseline=-3pt]
\path (0,0) circle [radius=1cm];
\draw (-2,0) -- (2,0);
\path[tips, -{Latex[scale length=1.5,scale width=1.125]}] (-2,0) -- (0,0);
\end{tikzpicture}
\end{equation}
and rearrange the perturbative expansion in terms of the self-energy (the sum over one-quark-irreducible graphs)
\begin{equation}\label{sigma}
\hat{\Sigma}\quad=\quad\begin{tikzpicture}[scale=0.2,line width=0.4pt,baseline=0.25cm]
\path (0,0) circle [radius=1cm] node[above={0.15cm}] {$\hat{\Sigma}$};
\draw (-5,0) arc [start angle=180, end angle=0, radius=5cm] -- cycle;
\end{tikzpicture}\quad,
\end{equation}
in the following way:
\begin{align*}
\begin{tikzpicture}[scale=0.45,line width=0.4pt,baseline=-3pt]
\draw (0,0) circle [radius=1cm];
\draw (-2,0) -- (-1,0) (1,0) -- (2,0);
\path (0,0) node {$\langle\hat{\mathcal{G}}\rangle$};
\path[tips, -{Latex[scale length=1.5,scale width=1.125]}] (-2,0) -- (-1.25,0);
\path[tips, -{Latex[scale length=1.5,scale width=1.125,reversed]}] (2,0) -- (1.25,0);
\end{tikzpicture}
\quad\!&=\!\quad
\begin{tikzpicture}[scale=0.25,line width=0.4pt,baseline=-3pt]
\path (0,0) circle [radius=1cm];
\draw (-2,0) -- (2,0);
\path[tips, -{Latex[scale length=1.5,scale width=1.125]}] (-2,0) -- (0,0);
\end{tikzpicture}
\quad\!+\!\quad
\begin{tikzpicture}[scale=0.45,line width=0.4pt,baseline=0.15cm]
\draw (-1.5,0) arc [start angle=180, end angle=0, radius=1.5cm] -- cycle;
\draw (-2.5,0) -- (-1.5,0) (1.5,0) -- (2.5,0);
\path[tips, -{Latex[scale length=1.5,scale width=1.125]}] (-2.5,0) -- (-1.75,0);
\path[tips, -{Latex[scale length=1.5,scale width=1.125,reversed]}] (2.5,0) -- (1.75,0);
\path (0,0) node[above={-0.05cm}] {$\hat{\Sigma}$};
\path (0,0) circle [radius=1cm];
\end{tikzpicture}&\!\!\!\!+\!\!\quad
\begin{tikzpicture}[scale=0.45,line width=0.4pt,baseline=0.15cm]
\draw (-1.5,0) arc [start angle=180, end angle=0, radius=1.5cm] -- cycle;
\draw (3,0) arc [start angle=180, end angle=0, radius=1.5cm] -- cycle;
\draw (-2.5,0) -- (-1.5,0) (1.5,0) -- (4,0) (6,0) -- (7,0) ;
\path[tips, -{Latex[scale length=1.5,scale width=1.125]}] (-2.5,0) -- (-1.75,0);
\path[tips, -{Latex[scale length=1.5,scale width=1.125]}] (1.5,0) -- (2.75,0);
\path[tips, -{Latex[scale length=1.5,scale width=1.125,reversed]}] (7,0) -- (6.25,0);
\path (0,0) node[above={-0.05cm}] {$\hat{\Sigma}$};
\path (4.5,0) node[above={-0.05cm}] {$\hat{\Sigma}$};
\path (0,0) circle [radius=1cm];
\end{tikzpicture}
\quad\!\!\!\!+\!\!\quad\ldots
\end{align*}
This rearrangement of the perturbation series (valid to all orders in the large-$N$ expansion) is the diagrammatic representation of the relation 
\begin{equation}\label{propagator}
\langle \hat{\mathcal{G}}\rangle=\frac{1}{\hat{\mathcal{G}}_0^{-1}-\hat{\Sigma}}
\end{equation}
between $\langle \hat{\mathcal{G}}\rangle$ and $\hat{\Sigma}$, which is equivalent to the definition of $\hat{\Sigma}$ as the sum over one-quark-irreducible graphs. 

As was explained above, this rearrangement is allowed because $\langle\hat{\mathcal{G}}\rangle$ is an analytic function of $\eta$, and therefore the corresponding Born series is convergent. This would not have been possible before the procedure of Hermitization.

We can express \cite{FZ, FZ1,JF} $\hat{\Sigma}$ in terms of the connected cumulants of the distribution of $\M{A}$ and the full propagator $\langle\hat{\mathcal{G}}\rangle$. Since $\M{A}$ is drawn from the GUE ensemble \eqref{eq:GUE}, there is only one connected cumulant, namely, the quadratic one
\begin{equation}\label{cumulant}
\langle A_{ij}A_{kl}\rangle_c = {1\over Nm^2}\delta_{il}\delta_{jk}\,.
\end{equation}
In the limit $N\rightarrow\infty$ only planar diagrams survive, and we thus arrive at
\begin{equation}\label{SE}
\begin{tikzpicture}[scale=0.3,line width=0.4pt,baseline=0.75cm]
\path (0,0) circle [radius=1cm] node[above={0.4cm}] {$\hat{\Sigma}$};
\draw (-5,0) arc [start angle=180, end angle=0, radius=5cm] -- cycle;
\end{tikzpicture}
\quad\!=\!\quad
\centering\begin{tikzpicture}[scale=0.35,line width=0.4pt,baseline=0.75cm]
\draw (0,0) circle [radius=1.2cm];
\draw (5.5,0) arc [start angle=0, end angle=180, radius=5.5cm];
\draw (-1,0) -- (-5,0) arc [start angle=180, end angle=0, radius=5cm] -- ((1,0);
\filldraw[fill=white] (0,5.2) circle [radius=1.2cm]; 
\path (0,0) node {$\langle\hat{\mathcal{G}}\rangle$};
\path (0,5.2) node { $\left\langle \! \mathcal{A}^2 \! \right\rangle_{\!\mathrm{c}}$};
\path[tips, -{Latex[scale length=2,scale width=1.5]}] (-5,0) -- (-2.8,0);
\path[tips, -{Latex[scale length=2,scale width=1.5,reversed]}] (5,0) -- (2.8,0);
\path[tips, -{Latex[scale length=2,scale width=1.5]}](-5.5,0) arc [start angle=180, end angle=135, radius=5.5cm];
\path[tips, -{Latex[scale length=2,scale width=1.5,reversed]}](-5,0) arc [start angle=180, end angle=135, radius=5cm];
\path[tips, -{Latex[scale length=2,scale width=1.5,reversed]}](5.5,0) arc [start angle=0, end angle=45, radius=5.5cm];
\path[tips, -{Latex[scale length=2,scale width=1.5]}](5,0) arc [start angle=0, end angle=45, radius=5cm];
\end{tikzpicture}
\quad\!=\!\quad \M{1}_N\otimes {1\over m^2}\left(\begin{array}{cccc} 
\overline{44}&0&0&\overline{41}\\
0&0&0&0\\
0&0&0&0\\
\overline{14}&0&0&\overline{11}
\end{array}\right).
\end{equation}
Here 
\begin{equation}\label{blocktrace}
\overline{\alpha \beta} = \frac{1}{N}  \langle \tr \hat{\mathcal{G}}_{\alpha\beta}\rangle\,,\quad \alpha, \beta :=1,2,3,4       
\end{equation}
are averaged traces over the $N\times N$ blocks $\langle\hat{\mathcal{G}}_{\alpha\beta}\rangle$ introduced in \eqref{blocks}. The diagrams in \eqref{SE} depict Eqs.~\eqref{fluctA} and \eqref{SE-def} in Appendix \ref{SE-details}, where one can find a detailed derivation of \eqref{SE} and other subsequent results. 

The block traces \eqref{blocktrace} are scalars, leading to the $4\times 4$ matrix factor on the right-hand side of Eq.~\eqref{SE}. Each block of $\hat\Sigma$ is proportional to the $N$-dimensional unit matrix $\M{1}_N$ after averaging over the unitary-invariant ensemble \eqref{eq:A}. The zero-rich block texture of $\hat\Sigma$ is very helpful practically, and it arises due to the decoupling of the random blocks from the deterministic blocks, which was the motivation for introducing \eqref{H2N} in the first place. Thus, $\hat\Sigma$ is determined exclusively by the four block traces $\overline{44}$, $\overline{41}$, $\overline{14}$ and $\overline{11}$.

Substitution of \eqref{SE} back in \eqref{propagator} leads to the so-called gap equation
\begin{equation}\label{gap}
\left\langle\left(\begin{array}{cccc}
\hat{\mathcal{G}}_{11}&\hat{\mathcal{G}}_{12}&\hat{\mathcal{G}}_{13}&\hat{\mathcal{G}}_{14}\\
\hat{\mathcal{G}}_{21}&\hat{\mathcal{G}}_{22}&\hat{\mathcal{G}}_{23}&\hat{\mathcal{G}}_{24}\\
\hat{\mathcal{G}}_{31}&\hat{\mathcal{G}}_{32}&\hat{\mathcal{G}}_{33}&\hat{\mathcal{G}}_{34}\\
\hat{\mathcal{G}}_{41}&\hat{\mathcal{G}}_{42}&\hat{\mathcal{G}}_{43}&\hat{\mathcal{G}}_{44}\\
\end{array}\right)\right\rangle = \left(\begin{array}{cccc} 
\eta-{\overline{44}\over m^2}&0&z&-{\overline{41}\over m^2}\\
0&\eta&-\M{B}&z\\
z^*&-\M{B}&\eta&0\\
-{\overline{14}\over m^2}&z^*&0&\eta-{\overline{11}\over m^2}
\end{array}\right)^{-1}, 
\end{equation} 
which determines all the blocks $\langle\hat{\mathcal{G}}_{\alpha\beta}(\eta;z,z^*)\rangle$ explicitly. (Note that in the matrix to be inverted on the right hand side of \eqref{gap}, we have suppressed explicit factors of the unit matrix $\M{1}_N$ in the appropriate blocks.)

This substitution of \eqref{SE} back in \eqref{propagator} amounts to resummation of the perturbation series. Thus, at this stage of solving the gap equation, we can safely set $\eta=is\rightarrow 0$ everywhere. Detailed subsequent calculations of the gap equation \eqref{gap} at $\eta=0$ are given in Appendix \ref{gap-details}.

As we can see from \eqref{etas}, in the limit $s\rightarrow 0$, 
\begin{eqnarray}\label{limitsOD}
\left\langle\left(\begin{array}{cc} \hat{\mathcal{G}}_{31} & \hat{\mathcal{G}}_{32} \\
\hat{\mathcal{G}}_{41} & \hat{\mathcal{G}}_{42}\end{array}\right)\right\rangle_{|_{s=0}} &=& 
\left\langle\left(\begin{array}{cc} 
\frac{z}{z^2-\M{A}\M{B}}& \M{A}\frac{1}{z^2-\M{B}\M{A}}\\
\M{B}\frac{1}{z^2-\M{A}\M{B}}& \frac{z}{z^2-\M{B}\M{A}}
\end{array}\right)\right\rangle = \left\langle{1\over z-\M{H}}\right\rangle\nonumber\\{}\nonumber\\
\left\langle\left(\begin{array}{cc} \hat{\mathcal{G}}_{13} & \hat{\mathcal{G}}_{14} \\
\hat{\mathcal{G}}_{23} & \hat{\mathcal{G}}_{24}\end{array}\right)\right\rangle_{|_{s=0}} &=& \left\langle\left(\begin{array}{cc} \hat{\mathcal{G}}_{31} & \hat{\mathcal{G}}_{32} \\
\hat{\mathcal{G}}_{41} & \hat{\mathcal{G}}_{42}\end{array}\right)\right\rangle^\dagger_{|_{s=0}}
\end{eqnarray}
and\footnote{The adjoint of the average is the average of the adjoint, because the probability distribution \eqref{eq:GUE} is real, of course.}
\begin{eqnarray}\label{limitsD}
\left\langle\left(\begin{array}{cc} \hat{\mathcal{G}}_{11} & \hat{\mathcal{G}}_{12} \\
\hat{\mathcal{G}}_{21} & \hat{\mathcal{G}}_{22}\end{array}\right)\right\rangle^\dagger_{|_{s=0}} &=&  - \left\langle\left(\begin{array}{cc} \hat{\mathcal{G}}_{11} & \hat{\mathcal{G}}_{12} \\
\hat{\mathcal{G}}_{21} & \hat{\mathcal{G}}_{22}\end{array}\right)\right\rangle_{|_{s=0}}
\nonumber\\{}\nonumber\\
\left\langle\left(\begin{array}{cc} \hat{\mathcal{G}}_{33} & \hat{\mathcal{G}}_{34} \\
\hat{\mathcal{G}}_{43} & \hat{\mathcal{G}}_{44}\end{array}\right)\right\rangle^\dagger_{|_{s=0}} &=& - \left\langle\left(\begin{array}{cc} \hat{\mathcal{G}}_{33} & \hat{\mathcal{G}}_{34} \\
\hat{\mathcal{G}}_{43} & \hat{\mathcal{G}}_{44}\end{array}\right)\right\rangle_{|_{s=0}}\,.
\end{eqnarray}

In particular, we see from \eqref{limitsOD} and \eqref{limitsD} that 
$$\langle\hat{\mathcal{G}}_{11}(s=0)\rangle^\dagger =-\langle\hat{\mathcal{G}}_{11}(s=0)\rangle\,,\quad \langle\hat{\mathcal{G}}_{44}(s=0)\rangle^\dagger=-\langle\hat{\mathcal{G}}_{44}(s=0)\rangle $$
and
\begin{equation}\label{1144hat}
\langle\hat{\mathcal{G}}_{14}(s=0)\rangle = \langle\hat{\mathcal{G}}_{41}(s=0)\rangle^\dagger\,.
\end{equation}
Thus, the corresponding block traces, evaluated at $s=0$, must further satisfy
\begin{equation}\label{1144}
\overline{14} = \overline{41}^*\,,\quad \Re \,\overline{11} = \Re\, \overline{44} = 0\,.
\end{equation} 

For convenience, let us introduce the notations
\begin{equation}\label{abc}
a(z,z^*)  = -{1\over m^2} \overline{44},\,\, b(z,z^*) = -{1\over m^2} \overline{41}\quad{\rm and}\quad c(z,z^*) = -{1\over m^2} \overline{11}
\end{equation}
where all quantities are evaluated at $s=0$. Thus, $a$ and $c$ are pure imaginary and $\overline{14}  = -m^2 b^*$. Moreover, as was mentioned following \eqref{equal-sums} in Appendix \ref{sec:diagonals}, the imaginary quantities $a$ and $c$ have the same sign, and thus 
\begin{equation}\label{ac}
ac\leq 0.
\end{equation}

As can be seen from \eqref{14gap}  in Appendix \ref{gap-details}, which we copy here
\begin{equation}\label{14gap-main}
\left\langle\left(\begin{array}{cc}
\hat{\mathcal{G}}_{11}&\hat{\mathcal{G}}_{14}\\
\hat{\mathcal{G}}_{41}&\hat{\mathcal{G}}_{44}\\\end{array}\right)\right\rangle  =  {1\over ac - |b+z^2 \M{B}^{-1}|^2} \left(\begin{array}{cc} 
c& - (b+z^2 \M{B}^{-1})\\
-(b^*+z^{*2} \M{B}^{-1})&a\end{array}\right),
\end{equation}
a  subset of the gap equations determines the four block traces $\overline{11}$, $\overline{14}$, $\overline{41}$ and $\overline{44}$ self-consistently, which in turn determine all the remaining blocks of $\langle \hat{\mathcal{G}}\rangle$.  Here and in what follows, in order to avoid cluttering of our equations, we have defined (with some minimal abuse of matricial notations)
\begin{equation}\label{abbreviation}
|b+z^2 \M{B}^{-1}|^2=(b+z^2 \M{B}^{-1})(b^*+z^{*2} \M{B}^{-1}).
\end{equation}
As was discussed in Appendix \ref{gap-details}, by equating block traces on both sides of \eqref{14gap-main} and with the definition \eqref{abc} in mind, we obtain the set of self-consistent equations for $\overline{11}$, $\overline{14}$, $\overline{41}$ and $\overline{44}$, or equivalently,  the quantities $a, b$ and $c$, as 
\begin{eqnarray}\label{abc-selfconsistent-main}
a &=& -{a\over Nm^2} \tr  {1\over ac - |b+z^2 \M{B}^{-1}|^2} \nonumber\\{}\nonumber\\
c &=& -{c\over Nm^2} \tr  {1\over ac - |b+z^2 \M{B}^{-1}|^2} \nonumber\\{}\nonumber\\
b &=& {1\over Nm^2} \tr  {b^* + z^{*2}\M{B}^{-1}\over ac - |b+z^2 \M{B}^{-1}|^2}.  
\end{eqnarray}
Equating block traces associated with $\langle\hat{\mathcal{G}}_{14}\rangle$ just produces the complex conjugate of the last equation in \eqref{abc-selfconsistent-main}, as it should. As further consistency check, one can easily verify that with purely imaginary $a$ and $c$ substituted in \eqref{XYWZ}, the averaged blocks $\langle\hat{\mathcal G}_{\alpha\beta}\rangle$ obtained by equating  \eqref{prop1} and  \eqref{prop2} fulfill all the hermitian conjugation consistency conditions in \eqref{limitsOD} and \eqref{limitsD}. 

The first two equations in \eqref{abc-selfconsistent-main} suggest that $a(z,z^*)$ and $c(z,z^*)$ be equal, a proposition we could not prove for the corresponding diagonal block traces $11(is; z,z^*)$ and $44(is;z,z^*)$ in Appendix \ref{sec:diagonals}, before taking averages. Let us now prove that 
\begin{equation}\label{AOC}
a(z,z^*) = c(z,z^*)\,,
\end{equation}
which completes the set of self-consistency equations. To this end, we first compute all four averaged diagonal traces from the expressions for $\langle\hat{\mathcal G}_{\alpha\alpha}\rangle$ given in \eqref{14gap-main} and \eqref{12gap} and obtain
\begin{eqnarray}\label{diagonal-block-traces}
\overline{11}(i0 ;z,z^*) &=& -{c\over N} \tr {1\over -ac + |b+z^2 \M{B}^{-1}|^2} \nonumber\\
\overline{22}(i0 ;z,z^*) &=& - {c\over N} \tr {|z|^2 \M{B}^{-2} \over -ac + |b+z^2 \M{B}^{-1}|^2} \nonumber\\
\overline{33}(i0 ;z,z^*) &=& - {a\over N} \tr {|z|^2 \M{B}^{-2} \over -ac + |b+z^2 \M{B}^{-1}|^2} \nonumber\\
\overline{44}(i0 ;z,z^*) &=&  -{a\over N} \tr {1\over -ac + |b+z^2 \M{B}^{-1}|^2} 
\end{eqnarray}
An important observation is that the two distinct matrices under the trace in these expressions are positive definite, due to \eqref{ac}. Thus, each of the traces appearing in \eqref{diagonal-block-traces} is positive. Compute now the two sums of traces appearing in \eqref{equal-sums}
\begin{eqnarray}\label{diagonal sums}
\overline{11}(i0 ;z,z^*) + \overline{22}(i0 ;z,z^*) &=& -{c\over N} \tr {\mathcal P} \nonumber\\
\overline{33}(i0 ;z,z^*) + \overline{44}(i0 ;z,z^*) &=& -{a\over N} \tr  {\mathcal P}
\end{eqnarray}
where\begin{equation}\label{P}
\mathcal{P} = {1 +|z|^2\M{B}^{-2}\over -ac +  |b+z^2 \M{B}^{-1}|^2} 
\end{equation}
is a positive matrix, rendering  $\tr {\mathcal P} >0$. According to \eqref{equal-sums}, the two sums on the LHS of both equations in \eqref{diagonal sums} are equal, thus proving \eqref{AOC}.  

Combining \eqref{AOC} and \eqref{trace-interrelation1}, which implies that $a(z,z^*) = c(z^*,z)$, we conclude also that 
\begin{equation}\label{Asym}
a(z,z^*) = a(z^*,z).
\end{equation}

As was mentioned following \eqref{equal-sums} and \eqref{SVDos1} in Appendix \ref{sec:diagonals}, the sums of diagonal block traces on the LHS of \eqref{diagonal sums} serve as order parameters indicating the location of the {\em two-dimensional} support $\widetilde{\mathcal D}\subset\CC$ of the density of eigenvalues of $\M{H}$ in the large-$N$ limit\cite{FZ}. After averaging over $\M{A}$, this role is clearly played by $a(z,z^*) = c(z,z^*)$, which vanish only outside the domain of eigenvalues $\widetilde{\mathcal D}$.

Finally, after solving the self-consistent equations for $a=c$ and $b$ explicitly, we can then substitute them into the block (cf \eqref{31-final-appendix}) 
\begin{equation}\label{31-final-main}
\langle\hat{\mathcal{G}}_{31}\rangle =  \left\langle\frac{z}{z^2-\M{A}\M{B}}\right\rangle=- {z\M{B}^{-1}(b^*+z^{*2} \M{B}^{-1})\over a^2 - |b+z^2 \M{B}^{-1}|^2},
 \end{equation}
and read off from it the resolvent of $\PHI = \M{A}\M{B}$. Thus, we obtain  the desired Green's function \eqref{eq:Green} as
\begin{equation}\label{Green1}
G(w,w^*)= -{1\over N}\tr {\M{B}^{-1}(b^*+w^* \M{B}^{-1})\over a^2 - |b+w \M{B}^{-1}|^2},\quad w=z^2.
\end{equation}

\section{Solution of the Gap Equations and the Density of Eigenvalues in the Complex Plane and on the Real Axis}\label{gap-dos}

Before delving into the technical details of computing the density of eigenvalues of the PH matrix $\PHI$ from \eqref{Green1}, let us get oriented by recalling the useful analogy between the formulation of two-dimensional electrostatics in the complex plane and the problem of determining the density of eigenvalues of non-hermitian random matrices in the complex plane \cite{electrostatics}. 

\subsection{The Two-Dimensional Electrostatic Analog of Spectra of Pseudo-Hermitian Matrices}\label{PH-Electrostatics} 
Consider an $N\times N$ non-hermitian matrix $\M{X}$ with eigenvalues $\lambda_1,\ldots, \lambda_N$, taken from some probability ensemble. In the large-$N$ limit,  these eigenvalues typically become dense on average\footnote{This may hold true even before averaging, due to the phenomenon of self-averaging, whereby the eigenvalues of a single realization of a large random matrix follow the large-$N$ averaged distribution of the ensemble to very high precision.} in a two-dimensional domain ${\mathcal D}$ in the complex plane, with continuous two-dimensional eigenvalue distribution 
\begin{equation}\label{rho2}
\rho^{(2)}(x,y) = \left\langle {1\over N} \sum_{i=1}^N\delta (x-\Re \lambda_i)\delta(y-\Im \lambda_i)\right\rangle.
\end{equation}
Thus, $\rho^{(2)}(x,y)$ is supported in the two-dimensional domain ${\mathcal D}$ in the complex plane\footnote{For the moment, we consider here generic non-hermitian matrices, but in the specific context of our discussion of the spectral support $\widetilde{\mathcal D}$ of the matrix $\M{H}$ in the previous section, one should think of ${\mathcal D}$ as the image of $\widetilde{\mathcal D}$ under the mapping $w=z^2$.} of $w=x+iy$, and the associated Green's function (analogous to \eqref{eq:Green}) is
\begin{equation}\label{GX}
G_2(w,w^*)=\left\langle \frac{1}{N} \tr {1\over w -\M{X}} \right\rangle = \int\limits_{\mathcal D}  \dd x'\dd y' {\rho^{(2)}(x',y')\over w-(x'+iy')}, 
\end{equation}
which one readily recognizes as the planar electric field generated by the charge density $\rho^{(2)}(x,y)$. By applying Gauss' law to the electric field \eqref{GX} we thus recover the charge density, namely, the eigenvalue density 
\begin{equation}\label{rhoC}
\rho^{(2)}(x,y)=\frac{1}{\pi}\frac{\partial}{\partial w^*}G_2(w,w^*),
\end{equation}
indicating that $G_2(w,w^*)$ is not a holomorphic function of $w$ in the spectral domain ${\mathcal D}$. Its domain of analyticity in $w$ is rather the complementary domain ${\mathcal D}^c = \CC/{\mathcal D}$  of the complex plane, which does not contain any eigenvalues. It also follows from \eqref{rhoC} that if $\rho^{(2)}(x,y)$ is bounded throughout the spectral domain ${\mathcal D}$, then $G_2(w,w^*)$ is continuous everywhere.

In contrast, the eigenvalues of hermitian matrices $\M{X} = \M{X}^\dagger$, with real eigenvalues $\lambda_i$, would typically condense along a one-dimensional segment (or several segments) $\sigma$ of the real axis $\RR$, giving rise to a continuous one-dimensional density 
\begin{equation}\label{rho1}
\rho^{(1)}(x) = \left\langle {1\over N} \sum_{i=1}^N\delta (x-\lambda_i)\right\rangle
\end{equation}
supported along $\sigma$, with Green's function 
\begin{equation}\label{GXherm}
G_1(w)=\left\langle \frac{1}{N} \tr {1\over w -\M{X}} \right\rangle = \int\limits_{\sigma}  \dd x' {\rho^{(1)}(x')\over w-x'}, 
\end{equation}
namely, the planar electric field generated by a charged one-dimensional wire $\sigma$ placed along the real axis. $G_1(w)$ is manifestly a holomorphic function of $w$ everywhere off the real axis, and it suffers a discontinuous jump across the real axis. The singular one-dimensional charge density $\rho^{(1)}(x)$ which causes this jump, can be recovered from this discontinuity by means\footnote{$\rho^{(1)}(x)$ can be also obtained, of course, by applying Gauss' law as in \eqref{rhoC}, because $\frac{1}{\pi}\frac{\partial}{\partial w^*}{1\over w-x'} = \delta(x-x')\delta(y)$.} of the well-known relation 
\begin{equation}\label{discontinuity}
\lim_{\epsilon \to 0^+} \left(G_1(x-i\epsilon) - G_1(x+i\epsilon)\right) = 2\pi i \rho^{(1)} (x).
\end{equation}
In contrast, as was mentioned above, $G_2(w,w^*)$ is continuous everywhere, and in particular, across the real axis. 

Here we come to the crux of our preliminary discussion: PH matrices are sort of a hybridization of hermitian and non-hermitian matrices, in the sense that their eigenvalues are either real, or come in complex-conjugate pairs. A large $N\times N$ PH matrix $\PHI$ would typically have  macroscopic amounts, that is, finite fractions of $N$, of both real and pairs of complex conjugate eigenvalues. In the large-$N$ limit, the complex pairs of eigenvalues would condense in a two-dimensional domain ${\mathcal D}$, which is {\emph symmetric} with respect to the real axis $\RR$, giving rise to a continuous two-dimensional distribution $\rho^{(2)}(x,y)$, while the real eigenvalues would condense along a segment $\sigma$ of the real axis, giving rise to a continuous one-dimensional distribution $\rho^{(1)}(x)$. These eigenvalue densities are normalized to the corresponding fractions of $N$ of eigenvalues they account for, that is, 
\begin{equation}\label{fractions}
 \int\limits_{\mathcal D}  \dd x\dd y \rho^{(2)}(x,y) = \nu,\quad  \int\limits_{\sigma}  \dd x \rho^{(1)}(x) = 1-\nu
\end{equation}
with $0\leq\nu\leq1$. 
The combination of these two-dimensional and one-dimensional distributions then gives rise to  the Green's function  \eqref{eq:Green} of our PH matrix $\PHI$, namely, 
\begin{equation}\label{GPH}
G_\text{PH}(w,w^*)=\left\langle \frac{1}{N} \tr {1\over w -\PHI} \right\rangle = \int\limits_{\mathcal D}  \dd x'\dd y' {\rho^{(2)}(x',y')\over w-(x'+iy')} +  \int\limits_{\sigma} \dd x' {\rho^{(1)}(x')\over w-x'}\,.
\end{equation}
Due to the reflection symmetry of the domain ${\mathcal D}$ with respect to the real axis, originating from isospectrality of $\PHI$ and $\PHI^\dagger$, we can see that 
\begin{equation}\label{conjugation}
G_\text{PH}^*(w,w^*) = G_\text{PH}(w^*,w).
\end{equation}
Moreover, it follows from \eqref{GPH} that $G_\text{PH}(w,w^*)$ is continuous everywhere, except in an infinitesimal sliver along the real axis, across which it jumps discontinuously as in \eqref{discontinuity}, from which, in combination with \eqref{conjugation}, we can determine the one-dimensional density component of \eqref{GPH} according to\footnote{The two dimensional integral in \eqref{GPH} does not contribute to the RHS of \eqref{1drho}, because in the limit taken, its integrand is a product of $\rho^{(2)}(x',y')$, which is even in $y'$,  and the imaginary part of the kernel, which is odd in $y'$.}   
\begin{equation}\label{1drho}
\rho^{(1)}(x) = {1\over \pi }\lim_{\epsilon\to 0^+}\Im G_\text{PH}(x-i\epsilon, x+i\epsilon).
\end{equation}
$G_\text{PH}(w,w^*)$ is an analytic function of $w$ in the complementary domain ${\mathcal D}^c$ (and off the real axis), and evidently, has asymptotic behavior
\begin{equation}\label{asymptotic}
G_\text{PH}(w,w^*) \sim {\nu + (1-\nu)\over w} = {1\over w}
\end{equation}
as $w\to\infty$. (This asymptotic regime obviously always belongs in ${\mathcal D}^c$ if the spectral domain ${\mathcal D}\cup\sigma$ of $\PHI$ is compact.)

Finally, by applying Gauss' law to \eqref{GPH}, we obtain the density of eigenvalues as 
\begin{equation}\label{total-density}
\frac{1}{\pi}\frac{\partial}{\partial w^*}G_\text{PH}(w,w^*)  = \rho^{(2)}(x,y) + \rho^{(1)}(x)\delta(y),
\end{equation}
with $\rho^{(2)}(x,y)$ supported throughout ${\mathcal D}$ and $\rho^{(1)}(x)$ supported along $\sigma$. Thus, $\rho^{(2)}(x,y)$ can be extracted from \eqref{total-density} simply by subtracting the singular piece proportional to $\delta (y)$. In the simpler case where ${\mathcal D}\cap \sigma = \emptyset$, we can avoid this subtraction by restricting $w$ to ${\mathcal D}$, and determine $\rho^{(1)}(x)$ directly from \eqref{1drho}. We shall discuss such a case as an example in Section \ref{sec:model B}.

\subsection{Solution of the Gap Equations and Phase Structure}\label{phase structure}
Let us now substitute $a=c$ from \eqref{AOC} and $w=z^2$ in \eqref{abc-selfconsistent-main}. Thus, we are instructed to solve the self-consistent gap equations
\begin{eqnarray}\label{ab}
a &=& -{a\over Nm^2} \tr  {1\over a^2 - |b+w \M{B}^{-1}|^2} \nonumber\\{}\nonumber\\
b &=& {1\over Nm^2} \tr  {b^* + w^*\M{B}^{-1}\over a^2 - |b+w \M{B}^{-1}|^2}  
\end{eqnarray}
(supplemented by the complex conjugated last equation) for the purely imaginary quantity $a(w,w^*)$, that is, 
\begin{equation}\label{a2}
a^2(w,w^*)\leq 0,
\end{equation}
and complex $b(w,w^*)$. 

In this section we shall analyze these equations assuming a generic invertible metric $\M{B}$. As a concrete example, in Section \ref{sec:model B} we shall outline the explicit full solution of these equations for a particular metric. 

These equations are algebraic polynomial equations for the unknown functions, and therefore have multiple roots, which typically become degenerate at branch point singularities in the complex-$w$ plane\footnote{The complex coordinate $w$ is the only variable parameter in \eqref{ab}, because the metric $\M{B}$ is held fixed in a given model.}. These roots should be then sewn together into unique single-valued continuous solutions for $a$ and $b$, defined globally throughout the properly cut complex $w$-plane. Note that $a$ and $b$ may certainly depend on both $w$ and $w^*$, and are therefore not globally holomorphic functions.  

At a given point $w$ (away from any branch point) and in some small neighborhood around it,  the ``physical" solution of \eqref{ab} should be clearly unique and continuous. 

From the first equation in \eqref{ab} we see that at a given point $w$, the solution for $a(w,w^*)$ is either $a(w,w^*)=0$ or $a(w,w^*)\neq 0$. On the basis of continuity, whichever of these two possibilities that holds at $w$, should also be valid in a neighborhood of that point. In fact, following the discussion below \eqref{Asym} in the previous section, $a(w,w^*)\neq 0$ throughout the {\em two-dimensional} spectral domain ${\mathcal D}$ of our averaged PH matrix $\PHI = \M{A}\M{B}$, and vanishes in the complementary domain\footnote{Except possibly for a benign removable discontinuity along $\sigma$, as discussed following \eqref{SVDos1} in Appendix \ref{sec:diagonals}, which has no practical effect on our solution.}$\,{\mathcal D}^c = \CC/{\mathcal D}$. Thus, $a(w,w^*)$  serves as a local order parameter indicating whether $w\in {\mathcal D}$ or not. 

For reasons that should become clear below, we shall refer to the solution with $a(w,w^*)\neq0$ throughout ${\mathcal D}$ as the {\em non-holomorphic} phase, and to the complementary solution with $a(w,w^*)=0$ throughout ${\mathcal D}^c$ as the {\em holomorphic} phase.

We have established the role of $a(w,w^*)$ as the order parameter telling apart the two phases. The other function $b(w,w^*)$ also has a physical interpretation, as sort of generalized 
self-energy. We argue as follows: Let us recall the resolvent 
\begin{equation}\label{PH-resolvent}
\frac{1}{z}\langle\hat{\mathcal{G}}_{31}\rangle = \langle\hat G\rangle =  \left\langle\frac{1}{w-\M{A}\M{B}}\right\rangle=- {\M{B}^{-1}(b^*+w^* \M{B}^{-1})\over a^2 - |b+w \M{B}^{-1}|^2}
 \end{equation}
of $\PHI$ from \eqref{31-final-main}. Thus, we can rewrite the second equation in \eqref{ab} as 
\begin{equation}\label{bSE}
-b  = {1\over N m^2} \tr \left\langle\M{B}\hat G\right\rangle. 
\end{equation}
In the simplest case $\M{B} = \M{1}_N$ (discussed in Section \ref{SQH} below), this equation reads 
\begin{equation}\label{bGUE}
-b(w) = {1\over m^2} G(w),
\end{equation} 
which we readily recognize from the diagrammatic equation \eqref{SE}, considered for the plain hermitian GUE matrices $\M{A}$ of \eqref{eq:GUE}, as the relation
\begin{equation}\label{SigmaGUE}
\Sigma_\text{GUE}(w)  = {1\over m^2} G_\text{GUE}(w)
\end{equation}
between the self-energy and the Green's function. Eq. \eqref{bSE} merely generalizes this relation to PH matrices, with arbitrary metric $\M{B}$.

\subsubsection{The Holomorphic Phase, $a=0$}\label{hp}
Outside the spectral domain ${\mathcal D}$, $a(w,w^*)=0$ identically. Thus, the second equation in \eqref{ab} simplifies into 
\begin{equation}\label{bh}
b(w) = -{1\over Nm^2} \tr {1\over b(w)+ w\M{B}^{-1}},\quad w\in {\mathcal D}^c. 
\end{equation}
Therefore, $b(w)$ is manifestly a holomorphic function of $w$ in  ${\mathcal D}^c$, as is $a(w)\equiv 0$. 

Alluding to the comments made below \eqref{a2}, Eq. \eqref{bh} is an algebraic polynomial equation for $b$, of degree $d+1$, where $d$ is the number of {\em distinct} eigenvalues of $\M{B}$. A subset of these roots, subjected to appropriate boundary conditions, such as continuity and asymptotic behavior as $w\rightarrow\infty$, should then be sewn together as branches of a single-valued holomorphic function  $b(w)$ defined throughout the complementary domain ${\mathcal D}^c$. 

This multibranched structure of $b(w)$ indicates that the complementary domain ${\mathcal D}^c$ need not necessarily be a simply connected, or even a connected set. 
However, if the spectral domain ${\mathcal D}$ of $\PHI$ is compact, ${\mathcal D}^c$ should clearly have a connected subset ${\mathcal D}^c_\infty $ which contains $w\rightarrow\infty$. Thus, by considering \eqref{bh} in the domain ${\mathcal D}^c_\infty $, we can read off the asymptotic behavior of $b(w)$, 
\begin{equation}\label{bhasympt}
b(w) \sim -{1\over m^2}{\left({1\over N}\tr\M{B}\right)\over w},\qquad (w\rightarrow \infty),
\end{equation}
assuming $\tr \M{B}\neq 0$. This asymptotic behavior is consistent, of course with \eqref{limitsOD} and the definition \eqref{abc}, which imply that  
\begin{equation}\label{bdef}
b = -{1\over Nm^2} \tr \left\langle \M{B} {1\over w-\PHI}\right\rangle.
\end{equation}

If, however, $\tr \M{B}=0$, we have to expand the RHS of \eqref{bh} to the next order in $b/w$, and obtain $$b(w)\sim \left({1\over Nm^2}\tr\M{B}^2\right) {b(w)\over w^2}, $$ 
which leads to a contradiction, unless $b(w)$ vanishes identically in a neighborhood of $w\rightarrow\infty$, and therefore must vanish identically throughout  ${\mathcal D}^c_\infty $:
\begin{equation}\label{zerobh}
b(w) =0\quad \forall w\in {\mathcal D}^c_\infty\quad {\rm if}  ~ \tr\M{B} = 0.
\end{equation}
Enforcing this result on the expansion of \eqref{bdef} in inverse powers of $w\in{\mathcal D}^c_\infty$, we come to the non-trivial conclusion, that if $\tr\M{B}=0$, then all moments (and in particular, all even moments)
\begin{equation}\label{vanishing-moments}
\langle \tr (\M{B}\PHI^n)\rangle  = \langle \tr (\M{B}(\M{A}\M{B})^n)\rangle  = 0,\quad n\geq 0.
\end{equation}

Having solved for $b(w)$, we then substitute it (together with $a(w) = 0$) in \eqref{31-final-main}, to obtain the resolvent 
\begin{equation}\label{reolvent-holomorphic}
\left\langle\frac{1}{w-\PHI}\right\rangle= {1\over b(w)\M{B}+w},
 \end{equation}
and thus, from \eqref{Green1}, the desired Green's function in the holomorphic phase 
\begin{equation}\label{Ghol}
G(w) = {1\over N}\tr  \left\langle\frac{1}{w-\PHI}\right\rangle= {1\over N}\tr {1\over b(w)\M{B}+w}
\end{equation}
throughout ${\mathcal D}^c$. Comparing this equation with the equation defining the self-energy $\Sigma_\text{herm}(w)$ for ensembles of hermitian matrices, 
\begin{equation}\label{SEhermitian}
G_\text{herm}(w) = {1\over w-\Sigma_\text{herm}(w)},
\end{equation}
the interpretation \eqref{bSE} of $-b$ as self-energy for PH matrices $\PHI$, alluded to in \eqref{bSE}, becomes manifest in the holomorphic phase:
\begin{equation}\label{self-energy}
G(w) =   {1\over N}\tr {1\over w - \M{\Sigma}_\PHI},\quad \M{\Sigma}_\PHI=-b(w)\M{B}.
\end{equation}

By multiplying the numerator and denominator under the trace in \eqref{Ghol} by $\M{B}^{-1}$ and then using \eqref{bh}, we can rewrite \eqref{Ghol} more compactly as 
\begin{equation}\label{Ghol1}
G(w) = {m^2 b^2(w)+ 1\over w}.
\end{equation}

The asymptotic behavior \eqref{bhasympt} of $b(w)$ implies that
\begin{equation}\label{Gholasympt} 
 G(w) \sim {1\over w}, \qquad (w\rightarrow \infty),
\end{equation}
in accordance with the definition of the resolvent of $\PHI$ and the assumed compactness of ${\mathcal D}$, and of course, with \eqref{asymptotic}. 

In the special case \eqref{zerobh} we have 
\begin{equation}\label{zerobh-Ghol}
G(w) = {1\over w}
\end{equation}
identically throughout ${\mathcal D}^c_\infty$.

Finally, the presence of a one-dimensional density component of real eigenvalues $\rho^{(1)}(x)$ in \eqref{GPH} means that $G(w)$ should have a cut along $\sigma\subset\RR$, with $\rho^{(1)}(x)$ given by the discontinuity across the cut as  
\begin{equation}\label{discontinuity1}
\lim_{\epsilon \to 0^+} \left(G(x-i\epsilon) - G(x+i\epsilon)\right) = 2\pi i \rho^{(1)} (x),
\end{equation}
in accordance with \eqref{discontinuity}.

\subsubsection{The Non-Holomorphic Phase, $a\neq0$}\label{nhp}
Throughout the spectral domain ${\mathcal D}$ the order parameter $a(w,w^*)\neq 0$ and is pure imaginary, and we can rewrite \eqref{ab} throughout this domain as
\begin{eqnarray}\label{ab-nh}
1 &=& {1\over Nm^2} \tr  {1\over |b+w \M{B}^{-1}|^2 + \alpha^2} \nonumber\\{}\nonumber\\
b &=&- {1\over Nm^2} \tr  {b^* + w^*\M{B}^{-1}\over  |b+w \M{B}^{-1}|^2 + \alpha^2}  
\end{eqnarray}
where 
\begin{equation}\label{alpha}
\alpha(w,w^*)  = \Im a(w,w^*).
\end{equation}
By using the first equation in \eqref{ab-nh} in the second one, we can rewrite the latter as 
\begin{equation*}
b = -b^* + {w^*\over Nm^2} \tr  {\M{B}^{-1}\over  |b+w \M{B}^{-1}|^2 + \alpha^2},   
\end{equation*}
that is, 
\begin{equation*}
2\Re b = {w^*\over Nm^2} \tr  {\M{B}^{-1}\over  |b+w \M{B}^{-1}|^2 + \alpha^2}, 
\end{equation*}
which leads to a contradiction, unless
\begin{eqnarray}\label{b-nh}
\hspace{-5cm}&&\Re b(w,w^*) \equiv 0,\quad \quad {\rm and}\nonumber\\{}\nonumber\\
&&\tr  {\M{B}^{-1}\over  |b+w \M{B}^{-1}|^2 + \alpha^2}  =0. 
\end{eqnarray}
Thus, $b(w,w^*)$ is pure imaginary throughout ${\mathcal D}$, like $a(w,w^*)$. These two functions are therefore manifestly not holomorphic functions of $w$ in ${\mathcal D}$. 

One can easily verify that the two real equations \eqref{b-nh}, together with the first equation in \eqref{ab-nh}, are completely equivalent to the original coupled equations 
\eqref{ab-nh}.Therefore, we can simplify \eqref{ab-nh} further, and rewrite them as two real equations for two unknown real functions as 
\begin{eqnarray}\label{ab-nh-real}
{1\over Nm^2} \tr  {1\over |i\beta+w \M{B}^{-1}|^2 + \alpha^2} &=& 1  \nonumber\\{}\nonumber\\
\tr  {\M{B}^{-1}\over  |i\beta +w \M{B}^{-1}|^2 + \alpha^2}  &=& 0,
\end{eqnarray}
where 
\begin{equation}\label{beta}
\beta(w,w^*)  = \Im b(w,w^*).
\end{equation}

Having solved for $a(w,w^*) = i\alpha(w,w^*)$ and $b(w,w^*) = i\beta(w,w^*)$, we then substitute them in \eqref{31-final-main}, to obtain the desired Green's function \eqref{eq:Green} in the non-holomorphic phase as
\begin{equation}\label{Green-nh}
G(w,w^*) =  {1\over N}\tr  \left\langle\frac{1}{w-\PHI}\right\rangle= {1\over N}\tr {\M{B}^{-1}(-i\beta+w^* \M{B}^{-1})\over  |i\beta+w \M{B}^{-1}|^2+\alpha^2}. 
\end{equation}
We can simplify this expression further into 
\begin{equation}\label{Green-nh-final}
G(w,w^*)= w^*\left[{1\over N}\tr { \M{B}^{-2}\over  |i\beta+w \M{B}^{-1}|^2+\alpha^2}\right],
\end{equation}
where we have used the second equation in \eqref{ab-nh-real}.  
Finally, we can compute the two-dimensional component $\rho^{(2)}(x,y)$ of the density of complex eigenvalues in the domain ${\mathcal D}$ by applying Gauss' law to \eqref{Green-nh-final} as in \eqref{total-density}.

\subsubsection{The Phase Boundary Line}\label{sec:boundary}
By definition, for probability ensembles with finite moments of the matrix elements of $\M{A}$, such as our Gaussian probability ensemble \eqref{eq:GUE}, the various averaged blocks $\langle\hat{\mathcal{G}}_{\alpha\beta}\rangle$ and their traces \eqref{blocktrace} should be continuous functions of $w$ and $w^*$. As we cross the phase boundary from the non-holomorphic phase into the holomorphic phase, $a(w,w^*)$ changes from being a non-zero pure imaginary function $i\alpha(w,w^*)$ into an identically vanishing function. This should happen continuously at the boundary line between the two phases,  and therefore the phase boundary line $\Gamma$ should be the solution of the equation 
\begin{equation}\label{boundary}
\alpha(w,w^*) = 0.
\end{equation}
That is, the phase boundary line $\Gamma$ is the line of zeros of $\alpha(w,w^*)$. 

Due to the structure of the gap equations \eqref{ab}, as we cross $\Gamma$ from the non-holomorphic phase into the holomorphic phase, $b$ would then change continuously from being a non-analytic pure imaginary function $i\beta(w,w^*)$ into a holomorphic function $b(w)$.  Let's assume for example that $\beta(w,w^*)$ is known explicitly in some subset of ${\mathcal D}$, which borders a subset ${\mathcal D}^c_1$ of the holomorphic domain ${\mathcal D}^c$ along an arc $\gamma\subset\Gamma$. In order to decide which branch of the solution of  \eqref{bh} we should assign to ${\mathcal D}^c_1$ we then use the fact that $b$ must be continuous across the phase boundary. Thus, our knowledge of $b(w,w^*) = i\beta(w,w^*)$ along the boundary arc $\gamma$ should fix the appropriate holomorphic branch of $b$ in a unique way. 

If it so happens that ${\mathcal D}^c_1$ forms a holomorphic island inside the non-holomorphic domain, so that $\gamma$ is actually a closed curve, then we can compute the holomorphic $b(w)$ inside ${\mathcal D}^c_1$ by invoking Cauchy's theorem as  
\begin{equation}\label{bhboundary}
b(w) = {1\over 2\pi i}\oint\limits_{\gamma} {i\beta(w',w^{'*})\over w-w'} \dd w'.
\end{equation}

\subsection{Positive Definite Metric $\M{B}$ : SQH Matrices}\label{SQH}
For positive definite metric $\M{B}$, our matrix $\PHI$ is SQH with purely real eigenvalues. It is easy to see that in this case, the gap equations admit only a holomorphic solution, resulting in purely real spectrum. Indeed, upon substituting  a positive definite  metric $\M{B}$ in the second equation in \eqref{ab-nh-real}, the matrix on the LHS of that equation becomes positive definite, with positive trace
\begin{equation}\label{ineq}
\tr  {\M{B}^{-1}\over  |i\beta +w \M{B}^{-1}|^2+ \alpha^2} >0.
\end{equation}
Thus, the non-holomorphic gap equation cannot be satisfied anywhere in the complex $w$-plane, and there can be no two-dimensional spectral domain accounting for complex eigenvalues. Only the holomorphic phase of Section \ref{hp} exists, with $a(w)\equiv 0$, $b(w)$ the appropriate solution of \eqref{bh}, and the holomorphic Green's function given by \eqref{Ghol}. The latter should have a cut along the support $\sigma$ of the condensed purely real eigenvalues of $\PHI$, and we should be able to recover their density from \eqref{discontinuity1}.

As a very simple check, let us verify that for $\M{B}=\M{1}_N$ our formalism reproduces Wigner's semicircle. From the interpretation of $-b$ as self-energy, in particular, from Equations \eqref{bSE} and \eqref{self-energy}, we know that 
\begin{equation}\label{bG}
-b(w) = {1\over m^2} G(w),
\end{equation} 
which is consistent, of course, with the definition \eqref{bdef} of $b$ in the case $\M{B} = \M{1}_N$ (and $\PHI = \M{A}$), namely
\begin{equation}\label{bdef1}
b = -{1\over Nm^2} \tr \left\langle  {1\over w-\M{A}}\right\rangle.
\end{equation}

Let us derive this relation explicitly from the holomorphic gap equation \eqref{bh}, which in the present case simplifies into the quadratic equation 
\begin{equation}\label{bh-GUE}
b(w) = -{1\over m^2}  {1\over b(w)+ w}, 
\end{equation}
from which it is also clear that we should pick the  root which behaves asymptotically like
\begin{equation}\label{bhasympt-GUE}
 b(w) \sim -{1\over m^2} {1\over w},\qquad (w\rightarrow \infty)
\end{equation}
in accordance with \eqref{bhasympt}. By comparing \eqref{bhasympt-GUE} with \eqref{Gholasympt}  we conclude \eqref{bG}, as required, because $b(w)$ is analytic in $w$ throughout the complex plane, save for a cut along the real axis. Thus, $-b(w) = \Sigma_\text{GUE}(w)$, which means that \eqref{Ghol} is simply the statement that 
\begin{equation}\label{Ghol:GUE}
G(w) = {1\over w + b(w)} = {1\over w-\Sigma_\text{GUE}(w)},
\end{equation}
that is, 
\begin{equation}\label{GGUE}
G(w) = {m^2\over 2}\left(w - \sqrt{w^2- {4\over m^2}}\right)
\end{equation}
is indeed the Green's function of the GUE ensemble, which leads to the semicircular eigenvalue density 
\begin{equation}\label{semicircle}
\rho^{(1)}(x)  = {m^2\over 2\pi} \sqrt{{4\over m^2} - x^2}
\end{equation}
supported in the segment $|x|\leq {2\over m}$, as required. 

\subsection{The Continuum Limit of the Density of Eigenvalues of the Metric $\M{B}$}\label{continuum}
As can be seen from Eqs. \eqref{bh} and \eqref{Ghol} in the holomorphic phase, and from Eqs. \eqref{ab-nh-real} and \eqref{Green-nh-final} in the non-holomorphic phase, all relevant quantities in these equations are {\em symmetric} functions of the eigenvalues $\mu_1, \mu_2,\ldots,\mu_N$ of the metric $\M{B}$. Thus, they must be all functionals of the density of eigenvalues 
\begin{equation}\label{Bdos}
\rho_B(\mu)  = \frac{1}{N}\sum_{k=1}^N \delta (\mu-\mu_k)
\end{equation}
of $\M{B}$, which generates all symmetric functions of these eigenvalues. This can be seen clearly by rewriting Eqs. \eqref{bh}, \eqref{Ghol}, \eqref{ab-nh-real} and \eqref{Green-nh-final}
explicitly in terms of \eqref{Bdos}, or more precisely, in terms of its Cauchy transform, the Green's function
\begin{equation}\label{GreenB}
G_B(w) = \frac{1}{N} \tr {1\over w-B}  = \int\limits_{-\infty}^{\infty} {\rho_B(\mu)\dd\mu\over w-\mu}
\end{equation}
of the metric. 

The main motivation for rewriting Eqs. \eqref{bh}, \eqref{Ghol}, \eqref{ab-nh-real} and \eqref{Green-nh-final} explicitly in terms of $G_B(w)$ is that in this form these equations are amenable to taking the continuum limit for $\rho_B(\mu)$. That is, the limit in which the $N\rightarrow\infty$ eigenvalues of $\M{B}$ condense in a finite segment (or segments) on the real axis, rendering $\rho_B(\mu)$ a continuous function in this domain\footnote{Let us assume a certain well-behaved continuous density function $\rho_B(\mu)$ as the continuum limit for the metric. Our working assumption is that as $N\rightarrow\infty$, the spectrum of the $N\times N$ hermitian matrix $\M{B}$ converges weakly to this given density $\rho_B(\mu)$.}. Yet another advantage of such reformulation of the gap equations is that it leads to general results and also to simplification of known results such as \eqref{invariant} and \eqref{G-nh-prefinal} below, which would be difficult to derive otherwise.

By definition \eqref{Bdos} is normalized according to 
\begin{equation}\label{Bdos-normalization}
\int\limits_{-\infty}^\infty \rho_B(\mu)\dd\mu = 1, 
\end{equation} 
resulting in the asymptotic behavior 
\begin{equation}\label{GreenB-asympt}
G_B(w) =  {1\over w} + {\langle\mu\rangle\over w^2}+{\langle\mu^2\rangle\over w^3}+\ldots
\end{equation}
as $w\rightarrow\infty$, where $$\langle\mu^k\rangle = \int\limits_{-\infty}^\infty \mu^k\rho_B(\mu) \dd\mu  = \frac{1}{N}\tr B^k$$ 
are the moments of $\rho_B(\mu)$. Note also the obvious reflection property 
\begin{equation}\label{GreenB-ref}
G_B(w^*) = G_B^*(w),
\end{equation}
which will be useful for us below.

\subsubsection{The Holomorphic Phase}\label{hol-sym}
In the holomorphic phase, we can clearly rewrite \eqref{bh} for determining $b(w)$ as 
\begin{eqnarray}\label{pre-bh-sym}
b(w) &=& -{1\over m^2} \int\limits_{-\infty}^\infty {\mu\rho_B(\mu)\over \mu b(w) + w}\dd\mu = -{1\over m^2 b(w)}\left[\int\limits_{-\infty}^\infty \left(1 - {w\over w+ \mu b(w)}\right)\rho_B(\mu)\dd\mu\right]\nonumber\\
&=&-{1\over m^2b(w)}\left[1 + {w\over b(w)}\int\limits_{-\infty}^\infty{\rho_B(\mu)\dd\mu\over -\frac{w}{b(w)} - \mu}\right],
\end{eqnarray}
where in the last step we have used \eqref{Bdos-normalization}. The integral in the last equation in \eqref{pre-bh-sym} can be expressed in terms of \eqref{GreenB}, and therefore can finally rewrite \eqref{bh} in terms of $G_B(w)$ as
\begin{equation}\label{bh-sym}
b(w) = -{1\over m^2 b(w)} \left[1 + {w\over b(w)} G_B\left(-\frac{w}{b(w)}\right)\right].
\end{equation}
It can be easily checked, based on \eqref{GreenB-asympt}, that \eqref{bh-sym} is consistent with the asymptotic condition \eqref{bhasympt}. 

To summarize, given the metric $\M{B}$, its Green's function $G_B(w)$ is known in principle, and therefore \eqref{bh-sym} constitutes a functional equation for determining $b(w)$.

Similarly, the Green's function $G(w)$ \eqref{Ghol} of $\PHI$ in the holomorphic phase can be written as 
\begin{equation}\label{Ghol-sym}
G(w) = -\frac{1}{b(w)}G_B\left(-\frac{w}{b(w)}\right),
\end{equation}
which means that the combination 
\begin{equation}\label{invariant}
wG(w) = -\frac{w}{b(w)}G_B\left(-\frac{w}{b(w)}\right)
\end{equation}
is invariant under the conformal mapping $w\mapsto -w/b(w)$. By substituting \eqref{invariant} in \eqref{bh-sym}  we obtain $m^2 b^2(w) + 1 = wG(w)$, which is nothing but \eqref{Ghol1}.

\subsubsection{The Non-Holomorphic Phase}\label{sec:nh-sym}
After multiplying the numerator and denominator of the coupled non-holomorphic gap equations \eqref{ab-nh-real} by $\M{B}^2$, we can rewrite them as  
\begin{eqnarray}\label{pre-nh-sym}
\int\limits_{-\infty}^\infty {\mu^2\rho_B(\mu) \dd\mu\over |i\beta\mu +w|^2 + \mu^2\alpha^2} &=&  \int\limits_{-\infty}^\infty {\mu^2\rho_B(\mu) \dd\mu\over x^2 + (y + \beta\mu)^2 + \mu^2\alpha^2} = m^2\nonumber\\
\int\limits_{-\infty}^\infty {\mu\rho_B(\mu) \dd\mu\over |i\beta\mu +w|^2 + \mu^2\alpha^2} &=& \int\limits_{-\infty}^\infty {\mu\rho_B(\mu) \dd\mu\over x^2 + (y + \beta\mu)^2 + \mu^2\alpha^2} = 0,
\end{eqnarray}
where $w=x+iy$, and with $\alpha(w,w^*)$ and $\beta(w,w^*)$ defined, respectively, in \eqref{alpha} and \eqref{beta}. 

Let us now multiply the first equation in \eqref{pre-nh-sym} on both sides by $\alpha^2 + \beta^2$. Then by adding and subtracting appropriate terms in its numerator, and after using the second equation in \eqref{pre-nh-sym} once, we can recast it into the form
\begin{eqnarray}\label{pre-nh-sym1}
m^2(\alpha^2 + \beta^2) &=& \int\limits_{-\infty}^\infty\left[1-{x^2 + y^2\over \alpha^2  + \beta^2}{1\over \left(\mu + {\beta y \over \alpha^2 + \beta^2}\right)^2 + \xi^2}\right]\rho_B(\mu)\dd\mu\nonumber\\
& =& 1 - {x^2 + y^2\over \alpha^2  + \beta^2} \int\limits_{-\infty}^\infty{\rho_B(\mu)\dd\mu\over \left(\mu + {\beta y \over \alpha^2 + \beta^2}\right)^2 + \xi^2},
\end{eqnarray}
where we have used \eqref{Bdos-normalization}, and where the real quantity $\xi$ is defined as 
\begin{equation}\label{xi}
\xi^2 = {\alpha^2(x^2 + y^2) + \beta^2 x^2\over (\alpha^2 + \beta^2)^2}.
\end{equation}
 Similarly, the second equation in \eqref{pre-nh-sym} is equivalent to 
 \begin{equation}\label{pre-nh-sym2}
\int\limits_{-\infty}^\infty{2\mu\rho_B(\mu)\dd\mu\over \left(\mu + {\beta y \over \alpha^2 + \beta^2}\right)^2 + \xi^2} = 0,
\end{equation}
which, like \eqref{ineq}, demonstrates that a solution to the non-holomorphic gap equations exists only if $\M{B}$ is not positive definite, that is, for such solutions to exist, $\rho_B(\mu)$ must be supported along both positive and negative segments. 

Let us now define the complex quantity 
\begin{equation}\label{zeta}
\zeta   =  i\xi - {\beta y\over \alpha^2 + \beta^2},
\end{equation}
in terms of which we can rewrite the denominator in \eqref{pre-nh-sym1} as
\begin{eqnarray}\label{den1}
\hspace{-0.5cm}{1\over (\mu +{\beta y\over \alpha^2 + \beta^2})^2 + \xi^2} &=& {1\over (\zeta - \mu)(\zeta^*-\mu)}\nonumber\\ &=& -\left({1\over \zeta-\mu} - {1\over \zeta^* - \mu} \right){1\over \zeta-\zeta^*} = {i\over 2\xi}\left({1\over \zeta -\mu} - {1\over \zeta^* - \mu} \right),
\end{eqnarray}
and the factor multiplying $\rho_B(\mu)$ in \eqref{pre-nh-sym2} as 
\begin{equation}\label{den2}
{2\mu\over (\mu +{\beta y\over \alpha^2 + \beta^2})^2 + \xi^2} = -\left({1\over \zeta-\mu} + {1\over \zeta^* - \mu} \right) + {\zeta + \zeta^*\over (\zeta - \mu)(\zeta^*-\mu)}.
\end{equation}
Thus, by plugging \eqref{den1} in \eqref{pre-nh-sym1} and \eqref{den2} in \eqref{pre-nh-sym2}, and then using \eqref{pre-nh-sym1} to calculate the integral over the last term in \eqref{den2}, we can rewrite the coupled non-holomorphic gap equations in terms of $G_B(w)$ as 
\begin{eqnarray}\label{nh-sym}
{i\over 2\xi}\left(G_B(\zeta) - G_B(\zeta^*)\right)  &=& {\alpha^2 + \beta^2\over x^2 + y^2}[1-m^2(\alpha^2 + \beta^2)]\nonumber\\
G_B(\zeta) + G_B(\zeta^*)  &=&  - {2\beta y \over x^2 + y^2}[1-m^2(\alpha^2 + \beta^2)].
\end{eqnarray}
 By recalling \eqref{GreenB-ref} we can simplify these two coupled real equations further into 
 \begin{eqnarray}\label{nh-sym-final}
\Im G_B(\zeta)   &=&- {\sqrt{\alpha^2 (x^2+y^2)  + \beta^2 x^2}\over x^2 + y^2}[1-m^2(\alpha^2 + \beta^2)]\nonumber\\
\Re G_B(\zeta)  &=&  - {\beta y \over x^2 + y^2}[1-m^2(\alpha^2 + \beta^2)], 
\end{eqnarray}
where we used the definition \eqref{xi} in the first equation. Alternatively, from \eqref{nh-sym}, we can represent these two real equations as a single complex gap equation 
\begin{equation}\label{nh-complex}
G_B(\zeta) = {\alpha^2 + \beta^2\over x^2 + y^2}\zeta^* [1-m^2(\alpha^2 + \beta^2)]
\end{equation}
or equivalently, 
\begin{equation}\label{nh-complex-final}
\zeta G_B(\zeta) = 1-m^2(\alpha^2 + \beta^2),
\end{equation}
where we used \eqref{xi} and \eqref{zeta}. Note that the RHS of this equation is purely real. 

Thus, to summarize, given the metric $\M{B}$, $G_B(w)$ is known in principle. The latter is evaluated at $\zeta$, which is a function of the non-holomorphic unknowns $\alpha$ and $\beta$. Therefore, the two coupled real gap equations \eqref{nh-sym-final}, or equivalently, the single complex gap equation \eqref{nh-complex-final},  determine $\alpha$ and $\beta$.

Once the non-holomorphic gap equations are solved for $\alpha$ and $\beta$, we could then use them to determine the resolvent $G(w,w^*)$ of $\PHI$ in the non-holomorphic phase. The latter is given by \eqref{Green-nh-final} as a trace over a function of the metric $\M{B}$. We can rewrite $G(w,w^*)$ in terms of $G_B(\zeta)$ by carrying algebraic manipulations similar to those carried above and obtain 
\begin{equation}\label{G-nh-1}
wG(w,w^*) = -{x^2+y^2\over \alpha^2 + \beta^2}{1\over\xi} \Im G_B(\zeta), 
\end{equation}
which by the first equation in \eqref{nh-sym} we can massage further into the remarkably simple form 
\begin{equation}\label{G-nh-prefinal}
wG(w,w^*)  = 1-m^2(\alpha^2 + \beta^2). 
\end{equation}
By comparing the latter equation with \eqref{nh-complex-final} we thus conclude the relation
\begin{equation}\label{invariant-nh}
wG(w,w^*)  = \zeta G_B(\zeta) =  1-m^2(\alpha^2 + \beta^2),
\end{equation}
which is the non-holomorphic counterpart of the holomorphic relations \eqref{invariant} and \eqref{Ghol1}.

\subsubsection{The Unified Form of the Gap Equation}
We can rewrite the holomorphic gap equation \eqref{bh-sym} as 
\begin{equation}\label{bh-sym-alt}
 - {w\over b(w)} G_B\left(-\frac{w}{b(w)}\right) = 1 + m^2 b^2(w).
 \end{equation}
Comparison of \eqref{bh-sym-alt} and its non-holomorphic counterpart \eqref{nh-complex-final} suggests that we extend $\zeta$, defined in the non-holomorphic phase by \eqref{zeta} and \eqref{xi}, into the entire complex $w$-plane in a natural way as
\begin{equation}\label{globalzeta}
\tilde\zeta = \left\{\begin{array}{cc} i\xi - {\beta y\over \alpha^2 + \beta^2}, ~~w\in {\mathcal D}\\{}\\- {w\over b(w)}, ~~w\in {\mathcal D}^c\end{array}\right.
\end{equation}
that is,  by setting $\alpha\mapsto 0$ and $\beta\mapsto -ib(w)$ in \eqref{xi} and \eqref{zeta}, so that $\zeta\mapsto -w/b(w)$ for $w\in{\mathcal D}^c$, rendering $\tilde\zeta$ continuous at the phase boundary. Then, the holomorphic gap equation \eqref{bh-sym-alt} (supplemented by $a\equiv 0$) and its non-holomorphic counterpart \eqref{nh-complex-final} could be unified into the single gap equation 
\begin{equation}\label{unified-gap}
\tilde\zeta G_B(\tilde\zeta) = 1 + m^2(a^2 + b^2),
\end{equation}
where it is understood that $a=0$ and $b(w)$ are holomorphic functions in the holomorphic regime $w\in {\mathcal D}^c$, and $a=i\alpha(w,w^*), b=i\beta(w,w^*)$ in the non-holomorphic regime $w\in {\mathcal D}$. 

In this way, once the gap equations are solved for the appropriate quantities $a$ and $b$, the two expressions \eqref{invariant} and \eqref{invariant-nh} for the Green's function of the PH  matrix $\PHI$ in the holomorphic and non-holomorphic regimes, respectively, are unified into the single form 
\begin{equation}\label{invariant-global}
wG(w,w^*)  =\tilde \zeta G_B(\tilde\zeta) =  1+m^2(a^2 + b^2).
\end{equation}
Thus, the combination $wG(w,w^*)  =\tilde \zeta G_B(\tilde\zeta)$ is invariant under the partly holomorphic, partly real-analytic mapping $w\mapsto\tilde\zeta$.

\subsubsection{The Phase Boundary Line $\Gamma$, Yet Again}
Let us see what our reformulation of the gap equations in terms of $G_B(w)$ entails for the behavior of our solutions at the phase boundary between the holomorphic and non-holomorphic phases discussed in Section \ref{sec:boundary}. Approaching the boundary from within the non-holomorphic phase we know that $\alpha\rightarrow 0$ as well as $b = i\beta$. Substituting these relations in \eqref{G-nh-prefinal} we immediately see that at the boundary 
\begin{equation}\label{bdry}
wG(w,w^*)  = 1 + m^2 b^2,  
\end{equation}
which is just \eqref{Ghol1} of the holomorphic phase, evaluated at the boundary. This should be expected, because the various holomorphic quantities should match continuously their non-holomorphic counterparts at the phase boundary. In this sense, \eqref{G-nh-prefinal} is quite natural, as it is a rather simple expression which crosses over continuously between the two phases at the phase boundary. 

We can derive \eqref{bdry} in yet another (but related) way, which underlines the continuity of $\tilde\zeta$ in \eqref{globalzeta} across the phase boundary line, as follows: At the boundary, 
\begin{equation}\label{xi-zeta}
\xi = \frac{x}{\beta},\quad \zeta = i{x+iy\over \beta} = {iw\over\beta},
\end{equation}
as can be seen from \eqref{xi} and \eqref{zeta}. By substituting \eqref{xi-zeta} in \eqref{nh-sym-final}, we can combine these two equations into
\begin{equation}\label{bdry1}
G_B\left(-\frac{w}{i\beta}\right) = -i\beta {1-m^2\beta^2\over w}, 
\end{equation}
in which both sides are evaluated at some point on the phase boundary line, where also $b = i\beta$. Thus, we can rewrite \eqref{bdry1}  as 
\begin{equation}\label{bdry2}
-{w\over b}G_B\left(-\frac{w}{b}\right) = 1+m^2 b^2, 
\end{equation}
which by \eqref{invariant}, evaluated at the phase boundary coming from the holomorphic phase, reproduces \eqref{bdry}. 

 \subsubsection{Some Checks and the Continuum Limit}
 In the latter part of Section \ref{SQH} we have verified that our formalism reproduced the GUE results for $\M{B} = \1_N$. For this metric, of course, $G_B(w) = \frac{1}{w-1}$, and one can easily verify that substituting this expression in \eqref{bh-sym} results in our previously derived quadratic equation \eqref{bh-GUE} for $b(w)$ in the GUE.  Similarly, substituting this $G_B(w)$ in \eqref{invariant} reproduces the Green's function \eqref{Ghol:GUE} of GUE. 
 
 A non-trivial check is offered by applying our results in this section to the metric \eqref{eq:metricB} studied in the next section, for which we readily write 
 \begin{equation}\label{GBB}
 G_B(w) = {k/N\over w-1} + {1-(k/N)\over w+1}.
 \end{equation}
We have verified that indeed, our results \eqref{bh-sym} and \eqref{invariant} in the holomorphic phase, and \eqref{nh-sym-final} and \eqref{G-nh-prefinal} in the non-holomorphic phase, reproduce all the analytic results reported in Section \ref{sec:model B}.

As was mentioned in the Introduction, in this paper we focus on invertible metrics $\M{B}$. That is, we only discuss in this paper such densities $\rho_B(\mu)$ which {\em vanish} in some neighborhood containing $\mu=0$. As a non-trivial example of such a {\em continuous} density, which is a simple generalization of \eqref{eq:metricB}, consider a metric whose positive and negative eigenvalues follow flat distributions along the positive and negative axis, namely, 
\begin{equation}\label{flat}
\rho_B(\mu)  = \left\{ \begin{array}{c} {1\over L_+ + L_-} \quad  -\mu_1 < \mu <-\mu_1 + L_-\\ 
\hspace{-0.3cm}{1\over L_+ + L_-} \quad\quad \mu_2-L_+ <\mu<\mu_2 \end{array}\right.
\end{equation}
for some $\mu_1>L_- >0$ and $\mu_2>L_+>0$. For this $\rho_B(\mu)$ we obtain 
\begin{eqnarray}\label{Gflat}
G_B(w) &=& {1\over L_+ + L_-}\left(\int\limits_{-\mu_1}^{-\mu_1+L_-} {\dd\mu\over w-\mu} + \int\limits_{\mu_2-L_+}^{\mu_2} {\dd\mu\over w-\mu}\right)\nonumber\\
 &=& {1\over L_+ + L_-}\log\left({w+\mu_1\over w+\mu_1-L_-}\cdot{w-\mu_2 + L_+\over w -\mu_2}\right), 
\end{eqnarray}
where the logarithms are defined in the cut plane, with a cut emanating from each branch point and running along the real axis in the negative direction. With this assignment of the cuts, the imaginary part of \eqref{Gflat} will have the correct discontinuity as in \eqref{discontinuity} along the support of \eqref{flat}. Substituting \eqref{Gflat} in \eqref{bh-sym},\eqref{invariant},\eqref{nh-sym-final} and \eqref{G-nh-prefinal} will clearly produce transcendental equations for the relevant quantities, which are not solvable analytically. We shall not pursue this problem any further in this paper.

\section{Example: The Density of Eigenvalues for\\ $\M{B}=\mathrm{diag}(1,\ldots,1,-1,\ldots,-1)$}\label{sec:model B}
In this Section we shall demonstrate the general formalism presented in the previous sections by means of a simple (but not too simple) example, corresponding to the indefinite diagonal metric 
\begin{equation}\label{eq:metricB}
\M{B}=\mathrm{diag}(\underbrace{1,\ldots,1}_{k},\underbrace{-1,\ldots,-1}_{N-k}),
\end{equation}
with $1<k<N$. 

Our discussion in this section of the metric \eqref{eq:metricB} will be somewhat telegraphic, avoiding detailed derivation of some of the analytical results and presenting only a limited set of numerical results.  We shall dedicate a subsequent paper \cite{FRuniversality} to providing the full details of the relevant analytical derivations, support them by results of ample numerical computations, including numerical demonstration of the statistical universality of this model, which is sensitive in the large-$N$ limit essentially only to the second cumulant \eqref{cumulant} of the random matrix $\M{A}$.

With the limit $N\rightarrow \infty$ in mind, let us define the fraction of the +1's on the diagonal of $\M{B}$ as
\begin{equation}\label{eq:lambda}
\lambda=\frac{k}{N}.
\end{equation}
From the general solution \eqref{eq:A} of the intertwining relation \eqref{intertwining}, and from the evenness symmetry of \eqref{eq:GUE},  we can see that the model has the obvious symmetry 
\begin{equation}\label{eq:symB}
\lambda \mapsto 1-\lambda.
\end{equation}

\subsection{The Holomorphic Phase, $a = 0$}\label{Bhol}
For the particular choice of metric $\M{B}$ given in \eqref{eq:metricB}, the holomorphic gap equation \eqref{bh} reduces to the cubic equation
\begin{equation}\label{eq:cubic}
m^2 b + \frac{\lambda}{b+w}+\frac{1-\lambda}{b-w}=0, 
\end{equation}
where according to \eqref{bhasympt}, we must pick that root of \eqref{eq:cubic} which behaves asymptotically as 
\begin{equation}
b(w)\sim \frac{1-2\lambda}{m^2 w},\qquad (w\rightarrow \infty).
\end{equation}

We can rewrite \eqref{eq:cubic} as
\begin{equation}
m^2b^3+(1-m^2w^2)b+w(1-2\lambda)=0. 
\label{eq:cubic2}
\end{equation}
Let us evaluate this cubic equation for $w=x+i0+$ with $x\in\RR$, i.e.~when $w$ approaches the real axis from above. Then for fixed $x$, the coefficients of the cubic equation \eqref{eq:cubic2} are real and its three possible solutions for $b$ are either all real or one of them is real and the other two come as a complex conjugated pair, depending on the sign of the discriminant $-\Delta/(27m^4),$ where
\begin{align}
\Delta&=\Delta(x)=\xi^2+4\cdot 27 m^6(1-m^2 x^2)^3,\nn\\
\xi&=\xi(x)=-27 m^4 (1-2\lambda)x.\label{eq:Deltaxi}
\end{align} 
We see from \eqref{eq:Deltaxi} that there are three real roots when $|x|$ is large. Furthermore, $\Delta(x)$ changes its sign at $x=\pm x_0,$ where
\begin{equation}
x_0= \Bigg( \tfrac{3|1-2\lambda|^{2/3}\,\big[\,\big(1-2\sqrt{\lambda(1- \lambda)} \big)^{1/3}+\big(1+2\sqrt{\lambda(1- \lambda)} \big)^{1/3}\,\big]+2}{2m^2}\Bigg)^{\!{}_{1/2}}. \label{eq:endpoints}\\
\end{equation}
We thus conclude that along the real axis \eqref{eq:cubic2} has one real and a pair of complex conjugate solutions for $b(x+i0+)$ when $x\in [-x_0,x_0],$ and three real solutions otherwise. 

For the particular metric \eqref{eq:metricB}, the general expression \eqref{Ghol} for the Green's function in the holomorphic phase yields
\begin{equation}\label{GholB}
G(w) = {\lambda\over b+w} - {1-\lambda\over b-w}.
\end{equation}
By substituting the appropriate root of \eqref{eq:cubic} in \eqref{GholB}, we can obtain the average density $\rho^{(1)}(x)$ of real eigenvalues of $\PHI$ from the discontinuity of this function across its cut along the real axis according to \eqref{discontinuity1} in the large $N$ limit. After a tedious but straightforward calculation, we find that $\rho^{(1)}(x)$ is supported along the interval $[-x_0,x_0]$ where it is given by
\begin{align}\label{eq:rhoreal}
\rho^{(1)}(x)&=\frac{1}{2\pi} \lim_{\epsilon\rightarrow 0+} \IM \Big[G(x- \ii \epsilon)-G(x + \ii \epsilon)\Big]\nn\\
&=\sign(1-2\lambda) \, \frac{\left|\xi-\sqrt{\Delta} \right|^{2/3} - \left|\xi+\sqrt{\Delta} \right|^{2/3}}{\sqrt{3} \cdot 2^{2/3}\cdot 6\pi m^2 x}, \qquad |x|<x_0,
\end{align}
where 
$\Delta$ and $\xi$ have been defined in \eqref{eq:Deltaxi} and $x_0$ in \eqref{eq:endpoints}.
\begin{figure}
\begin{center}
\includegraphics[width=\textwidth]{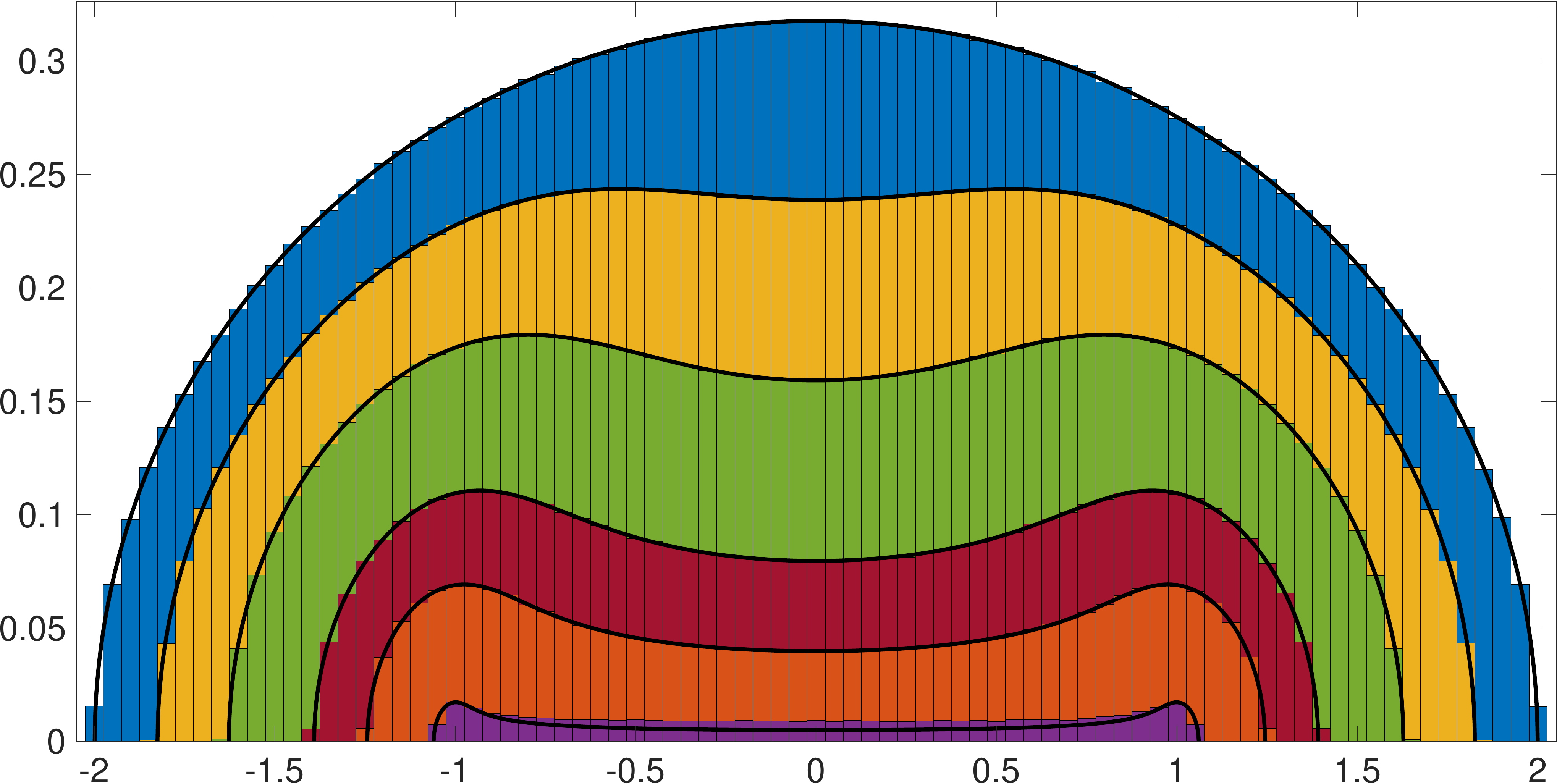} 
\end{center} 
\caption{Normalized histograms of real eigenvalues obtained from numerical simulation of 2000 samples of size $N=1024$ with $m=1$. Different colors represent different values of $\lambda$: $1/1024, 1/8, 1/4, 3/8, 7/16, 63/128$ (in order of decreasing density). The solid black lines show the corresponding large-$N$ theoretical predictions for $\rho^{(1)}(x)$ given by \eqref{eq:rhoreal}.}\label{fig:rhoreal}
\end{figure}
In particular, at the center of the band of real eigenvalues the density is given by
\begin{equation}
\lim_{x\rightarrow 0}\rho^{(1)}(x)=\frac{m|1-2\lambda|}{\pi}.
\end{equation}

In Fig.~\ref{fig:rhoreal} we show normalized histograms obtained from numerically generated samples for various values of $\lambda$. For each value of $\lambda$, we have used 2000 samples of matrix size $N=1024$. The histograms fit well with the solid black lines from the corresponding theoretical limits $N\rightarrow\infty$ given by Eq.~\eqref{eq:rhoreal}. A small deviation due to finite $N$ effect can be seen only in the histogram in magenta showing the result for $\lambda=63/128$, which is close to the degenerate case $\lambda = 1/2$, for which the theoretical density \eqref{eq:rhoreal} predicts that $\rho^{(1)}(x)\equiv 0$. As $\lambda$ decreases from $1/2$, the density of real eigenvalues increases. The blue histogram corresponds to having $\lambda$ close to zero. For $\lambda=0$ we obtain of course Wigner's semi-circular distribution, because then $\PHI=\M{A}$, which has been drawn from the GUE. 

According to \eqref{asymptotic}, the fractions of all real and all complex eigenvalues sum up to one. Integrating $\rho^{(1)}(x)$ yields the fraction $1-\nu$ of real eigenvalues as a function of $\lambda$, in accordance with \eqref{fractions}. Carrying out this integral explicitly is a rather difficult task, which we can avoid by the much easier calculation of the complementary fraction of complex eigenvalues in \eqref{eq:fracnonhol} in the next section \ref{Bnh}. Thus, we find that the fraction of real eigenvalues is 
\begin{equation} \label{eq:vshaped}
1-\nu=\int_{-x_0}^{x_0} \rho^{(1)}(x) \dd x= |1-2\lambda|,
\end{equation}
\begin{figure}
\begin{center}
\includegraphics[width=\textwidth]{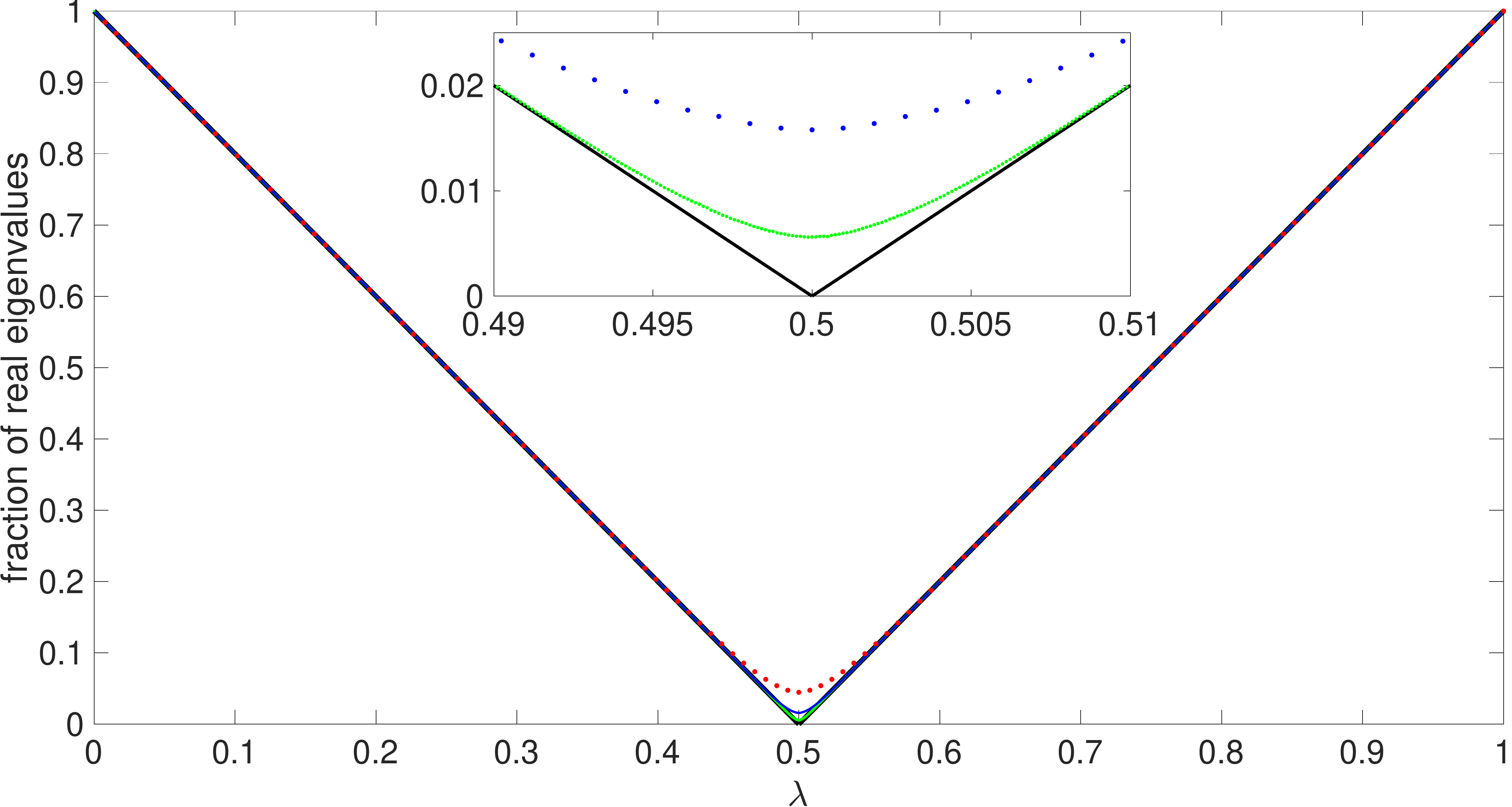}
\end{center} 
\caption{Fraction of real eigenvalues as a function of $\lambda$. Solid black line shows the theoretical large $N$ prediction given by \eqref{eq:vshaped}, while the dots represent numerical simulations with $N=128$ averaged over 500000 samples (red), $N=1024$ using 5000 samples (blue) and $N=8192$ using 10 samples (1000 samples for the inset) (green). Inset: Magnification near $\lambda=1/2$.
}\label{fig:vshaped}
\end{figure}
which is consistent with the symmetry  \eqref{eq:symB} of the model. In Fig.~\ref{fig:vshaped} this result has been compared with numerical simulations for matrix sizes $N=128$, $N=1024$ and $N=8192$ averaged over many samples. For $\lambda$ away from $1/2$ the convergence is exponentially fast. Near $1/2$, we can see a small deviation for finite $N$. The theoretical large $N$ prediction, actually is a lower bound for the number of real eigenvalues, even for finite $N$ and for each sample. This follows from a special case of a purely algebraic theorem \cite{carlson}.

\subsection{The Non-Holomorphic Phase, $a\neq 0$}\label{Bnh}
In this phase $a(w,w^*)=i\alpha(w,w^*)$ and $b(w,w^*) = i\beta(w,w^*)$ are pure imaginary. By substituting the metric $\M{B}$ given by \eqref{eq:metricB} in the gap equations \eqref{ab-nh}, or equivalently \eqref{ab-nh-real}, we thus obtain 
\begin{eqnarray}\label{ab-nh-real-B}
{1\over m^2} \left[ {\lambda\over \alpha^2 + |i\beta + w|^2}  +   {1-\lambda\over \alpha^2 + |i\beta - w|^2}\right]  &=& 1  \nonumber\\{}\nonumber\\
{\lambda\over \alpha^2 + |i\beta + w|^2}  -  {1-\lambda\over \alpha^2 + |i\beta - w|^2} &=& 0,
\end{eqnarray}
or equivalently
\begin{eqnarray}\label{ab-nh-real-B1}
\alpha^2 + |i\beta + w|^2  &=& {2\lambda\over m^2}  \nonumber\\{}\nonumber\\
\alpha^2 + |i\beta - w|^2  &=& {2(1-\lambda)\over m^2}.
\end{eqnarray}
Subtraction of these two equations yields a linear equation for $\beta$ from which we determine
\begin{equation}\label{beta-B}
\beta = {2\lambda-1\over 2m^2 y},
\end{equation}
where we substituted $w=x+iy$. Finally, feeding \eqref{beta-B} back in \eqref{ab-nh-real-B1} we determine
\begin{equation}\label{alpha-B}
\alpha^2 = {1\over m^2} - \left[x^2 + y^2 + \left({2\lambda-1\over 2m^2y}\right)^2\right]\geq 0. 
\end{equation}
\begin{figure}[b!]
\begin{center}
\includegraphics[width=0.495\textwidth]{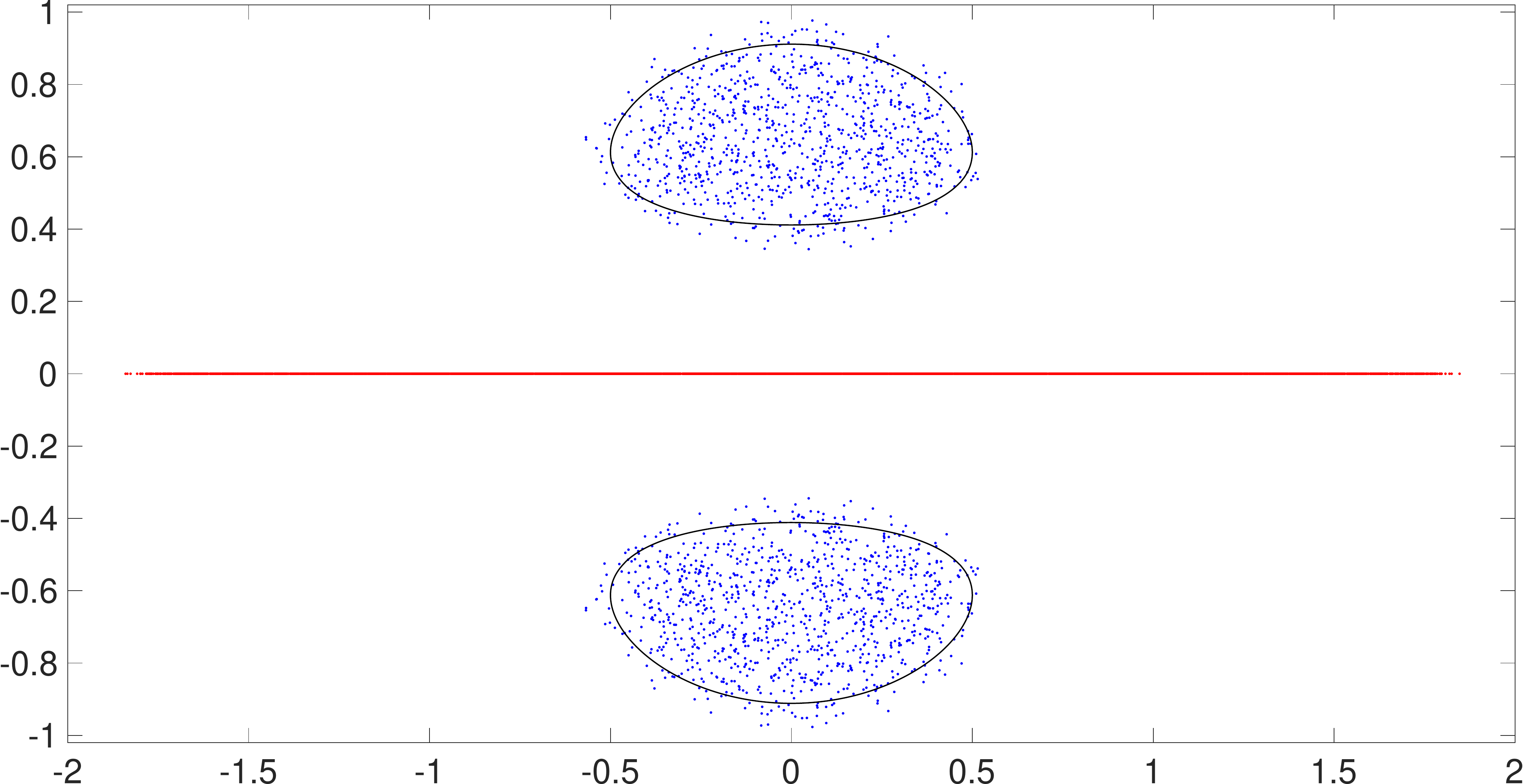}\hfill\includegraphics[width=0.495\textwidth]{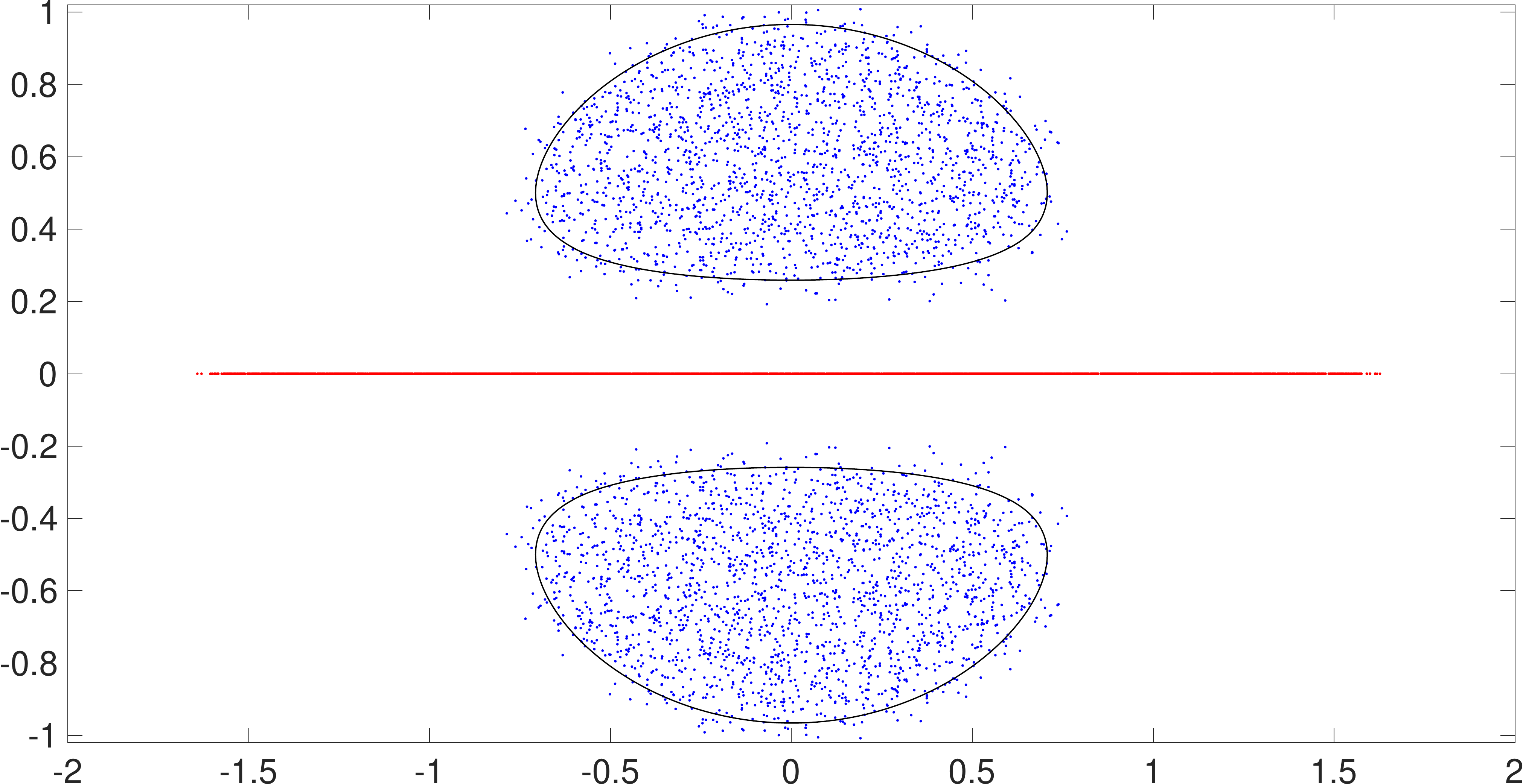}
\includegraphics[width=0.495\textwidth]{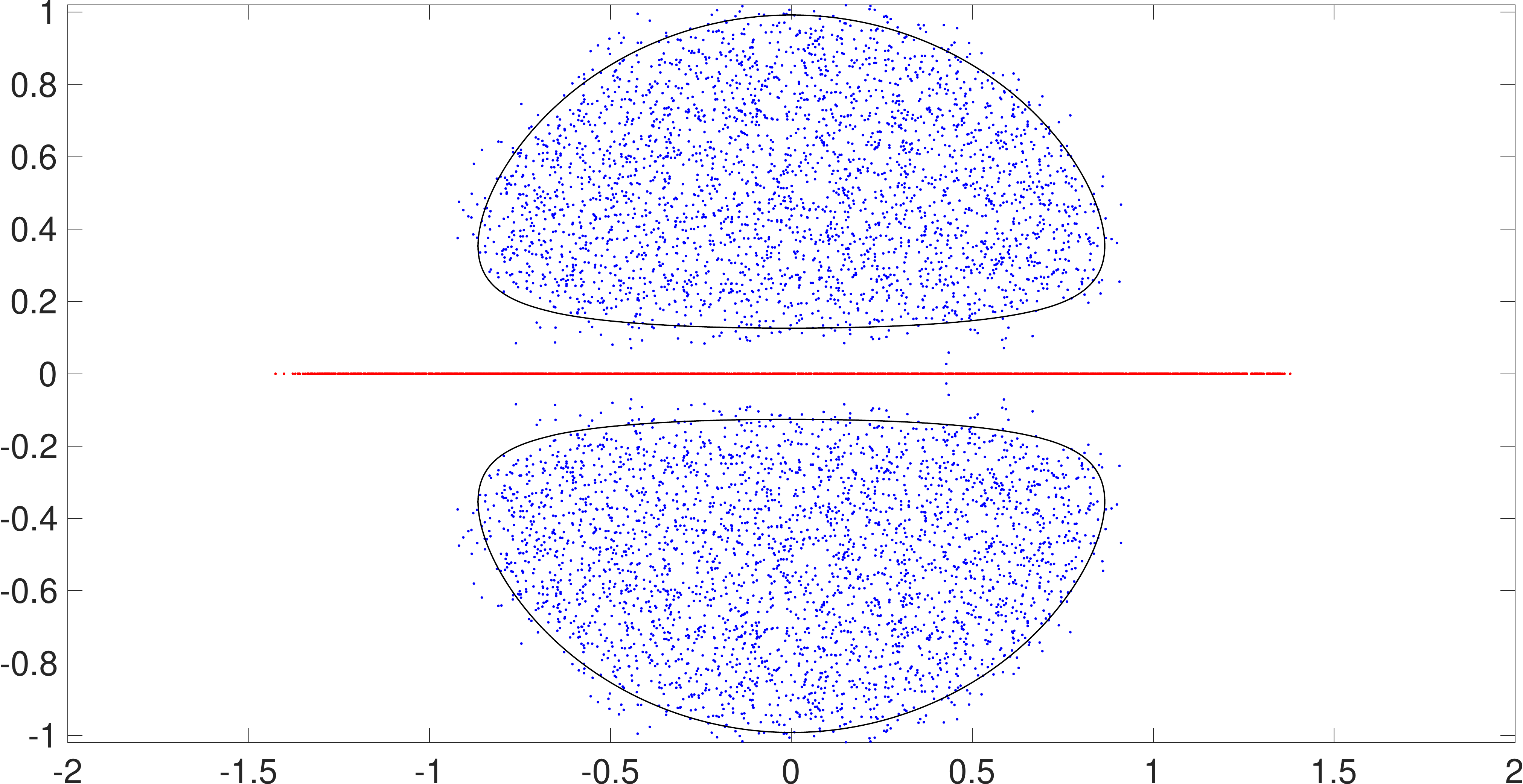}\hfill\includegraphics[width=0.495\textwidth]{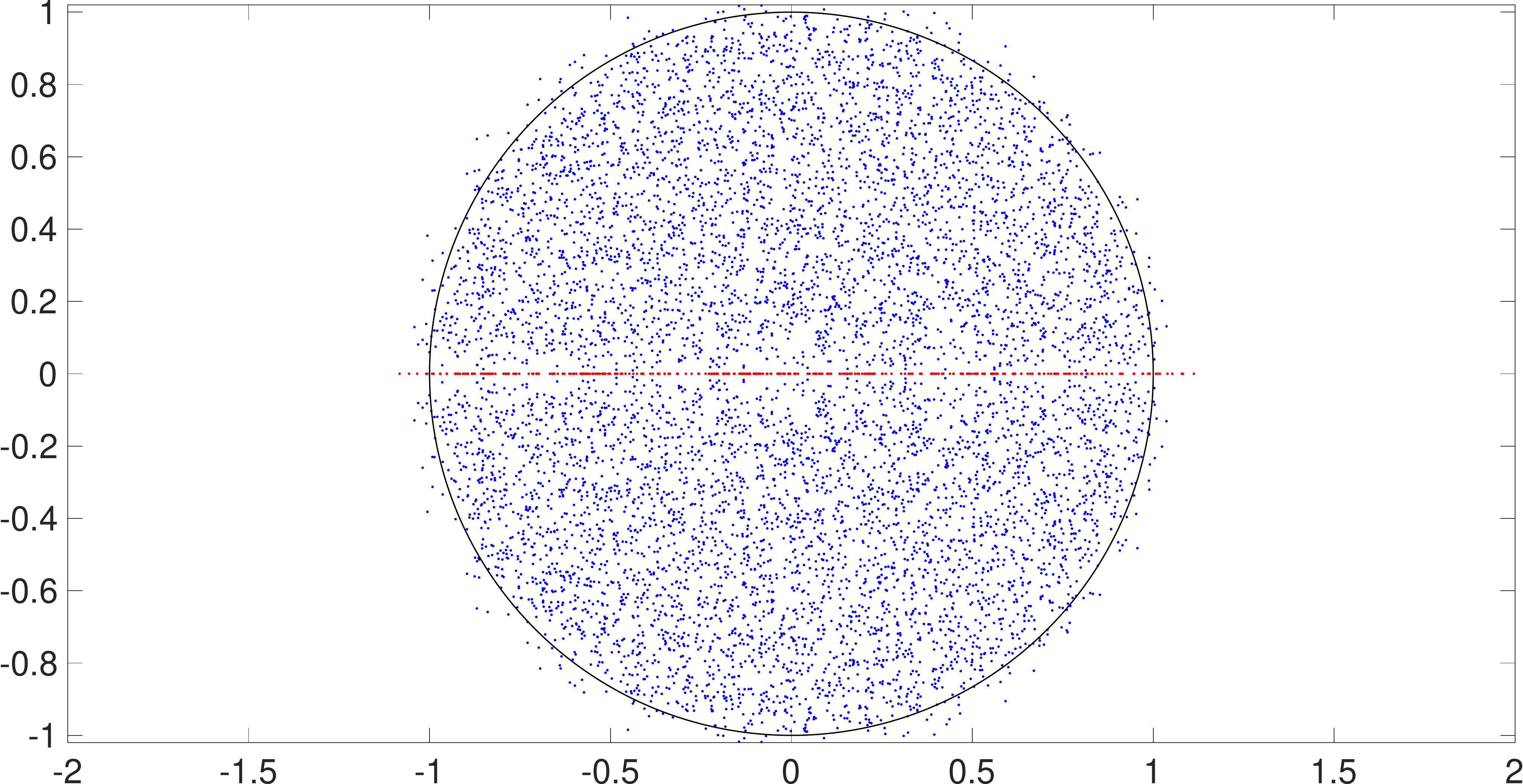} 
\end{center} 
\caption{Scatter plots showing the spectrum of $\phi$ for various values of $\lambda$, collected from 64 samples of matrix size $N=128$ with $m=1$. Complex eigenvalues are marked in blue, real ones in red. Solid black lines follow the theoretical $N\rightarrow\infty$ prediction of the phase boundary line in Eq.~\eqref{eq:rpm}.  Here $\lambda=1/8,1/4,3/8,1/2$ (upper left to lower right) }\label{fig:nonholomorphicN128}
\end{figure}
According to \eqref{boundary}, the boundary line between the holomorphic and non-holomorphic phases is the curve along which \eqref{alpha-B} vanishes, which is conveniently represented in polar coordinates $x=r\cos\theta, y=r\sin\theta$ by
\begin{equation}\label{boundary-B}
r^4\sin^2\theta - {r^2\sin^2\theta\over m^2} + \left({\sin\theta_0\over 2m^2}\right)^2=0
\end{equation}
where $\sin\theta_0 = |2\lambda-1|$. This curve is made of two closed loops, one in the upper half-plane, and its mirror image with respect to the real axis in the lower half-plane. Each loop is made of two arcs (the roots of \eqref{boundary-B}) 
\begin{equation}\label{eq:rpm}
r_\pm(\theta)=\frac{1}{\sqrt{2}m}\left( 1\pm \sqrt{1-\left( \frac{\sin \theta_0}{\sin \theta } \right)^2} \right)^{1/2}\!\!\!\!\!\!\!\!, 
\end{equation}
where $\sin^2\theta\ge \sin^2 \theta_0$. $r_+$ is the part of the boundary farther from the origin, and $r_-$ the closer one. The two arcs are sewn together at $\sin^2\theta = \sin^2\theta_0$ to make the loops.

These two loops enclose two two-dimensional blobs which comprise the domain ${\mathcal D}$ of the non-holomorphic phase that contains the condensed pairs of complex conjugate eigenvalues of $\PHI$ in the large-$N$ limit. 

We can also derive the boundary equation \eqref{boundary-B} coming from the holomorphic phase: As we approach the phase boundary curve, the holomorphic solution $b(w)$ of \eqref{eq:cubic2} should become purely imaginary. Thus, by substituting the boundary value $b=i\beta(x,y)$ in \eqref{eq:cubic2} and separating the resulting cubic into its real and imaginary parts, we obtain the two coupled equations 
\begin{eqnarray}\label{boundary-real-im}
m^2\beta^3 + [m^2(x^2-y^2)-1]\beta - (1-2\lambda)y &=& 0\nonumber\\
x[2m^2y\beta + 1-2\lambda] &=& 0.
\end{eqnarray}
By comparison with \eqref{beta-B}, we see from the second equation that the holomorphic $b(w)$ coincides with the non-holomorphic solution at the boundary, as it should, by continuity. Then, by substituting this boundary value of $\beta$ in the first equation in \eqref{boundary-real-im} we simply obtain the Cartesian form of \eqref{boundary-B} (multiplied by $2\lambda-1$).

The particular metric \eqref{eq:metricB} has the special property that $\M{B}^2=\mathbb{1}_N$, which upon substitution in \eqref{Green-nh-final} and use of the first gap equation in \eqref{ab-nh-real-B}, leads to the particularly simple non-holomorphic Green's function
\begin{equation}\label{G-nhol-B}
G(w,w^*) = m^2w^*,
\end{equation}
valid throughout ${\mathcal D}$. At the phase boundary line \eqref{eq:rpm} the non-holomorphic Green's function \eqref{G-nhol-B} has to cross-over continuously to the holomorphic Green's function \eqref{GholB}. 

Note in passing that \eqref{G-nhol-B} implies $wG(w,w^*) = m^2w^*w$. By comparing this with \eqref{G-nh-prefinal} we thus conclude that $\alpha^2 + \beta^2 = \frac{1}{m^2} - x^2-y^2$, which is consistent with \eqref{alpha-B}. 

By applying Gauss' law to \eqref{G-nhol-B}, as in \eqref{total-density}, we see that the density of complex eigenvalues
\begin{equation}\label{eq:rhoC}
\rho^{(2)}(x,y)=\frac{1}{\pi}\frac{\partial}{\partial w^*}G(w,w^*)=\frac{m^2}{\pi},
\end{equation}
is uniform in ${\mathcal D}$, and also independent of $\lambda$. 

The area of the two blobs of ${\mathcal D}$, enclosed by the two loops \eqref{eq:rpm}, is given by 
\begin{equation}\label{eq:area}
2 \int_{r_-}^{r_+} \int_{\theta_0-\pi}^{2\pi-\theta_0} r \dd r \dd \theta=  \int_{\theta_0-\pi}^{2\pi-\theta_0} \left(r_+^2(\theta)-r_-^2(\theta)\right)\dd \theta=\frac{(1-|1-2\lambda|)\pi}{m^2}.
\end{equation}
By multiplying this area and the density \eqref{eq:rhoC}, we find that the fraction $\nu$ of complex eigenvalue is given by
\begin{equation}\label{eq:fracnonhol}
\nu=1-|1-2\lambda|
\end{equation}
	
\begin{figure}[b!]
	\begin{center}
		\includegraphics[width=0.495\textwidth]{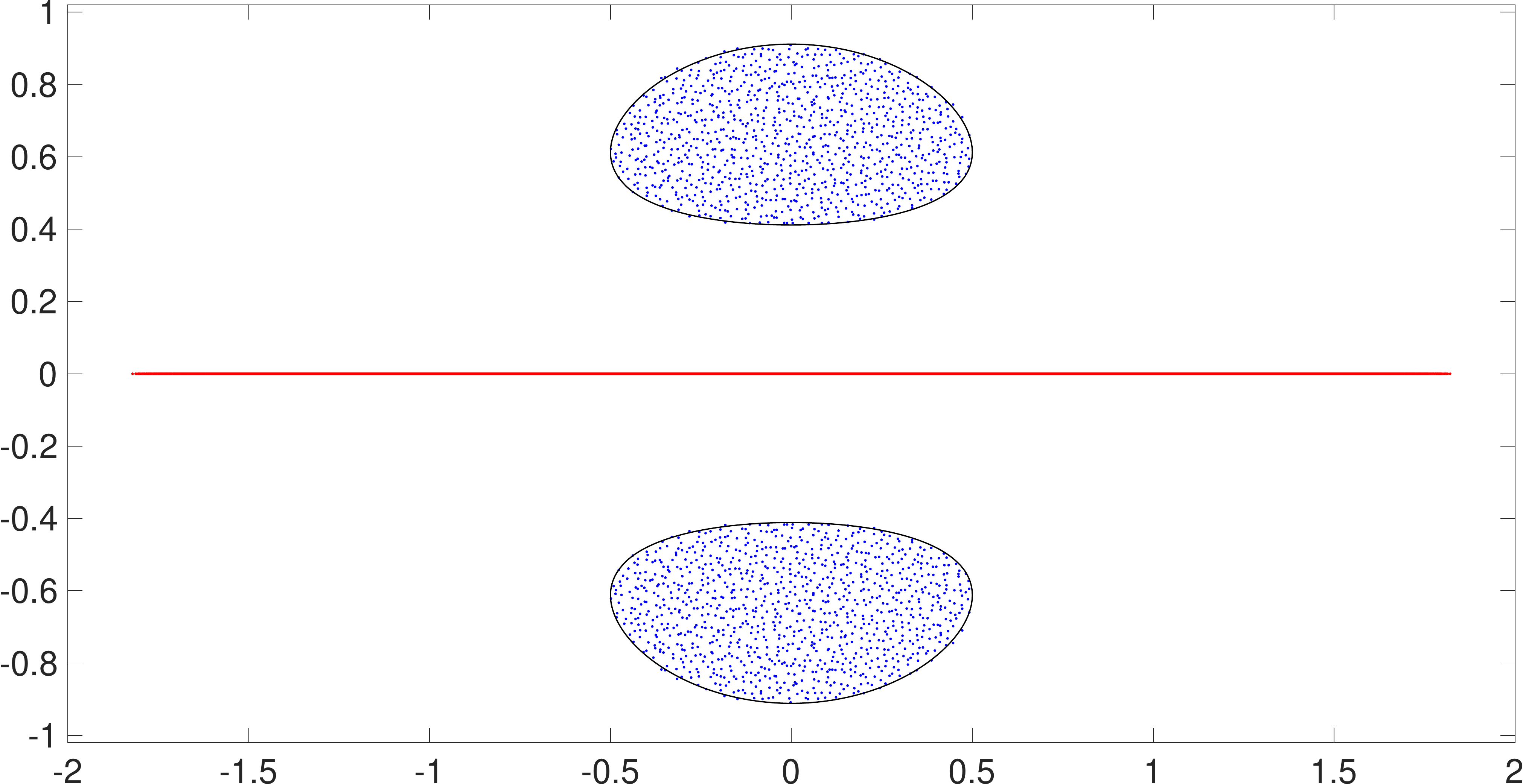}\hfill\includegraphics[width=0.495\textwidth]{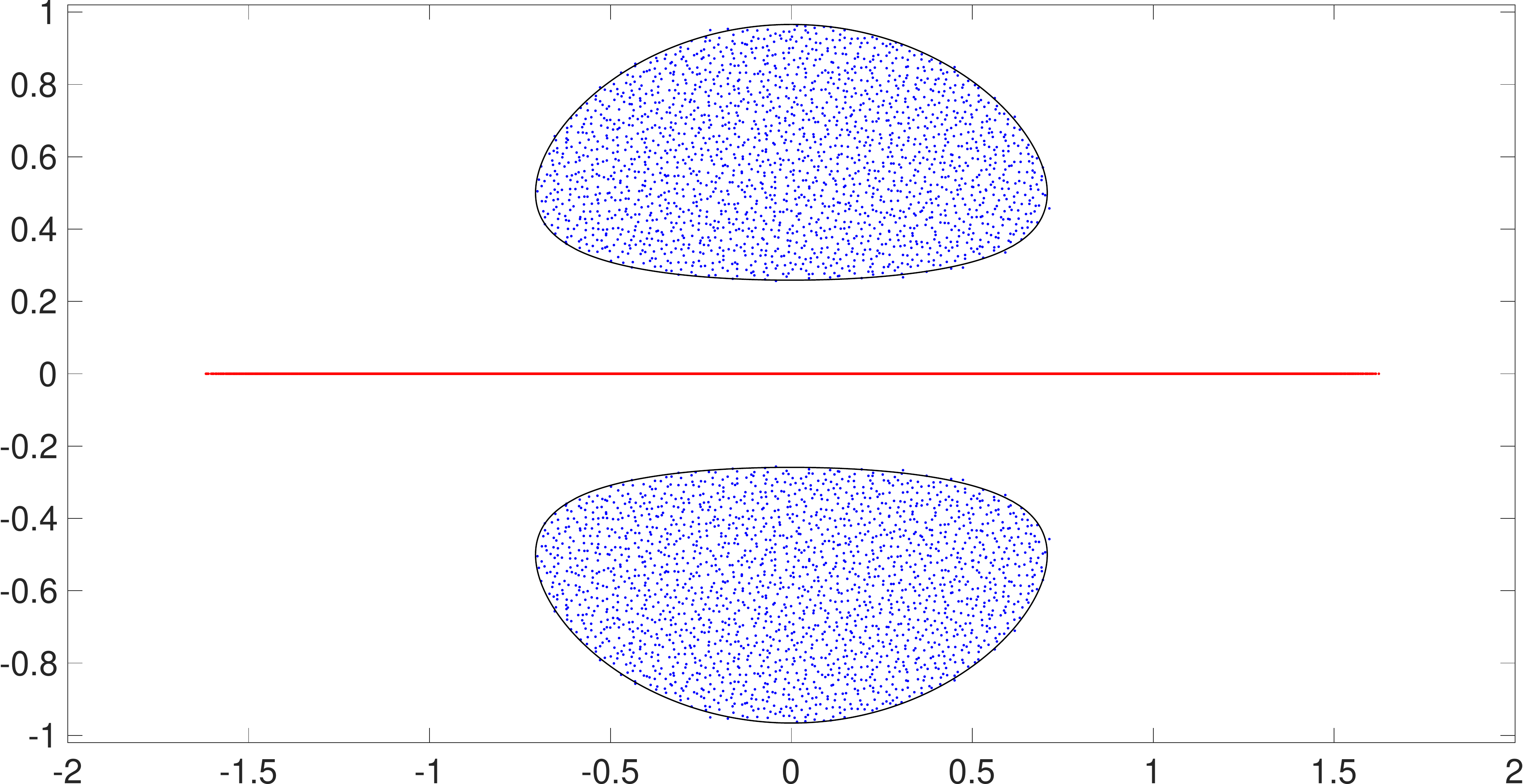}
		\includegraphics[width=0.495\textwidth]{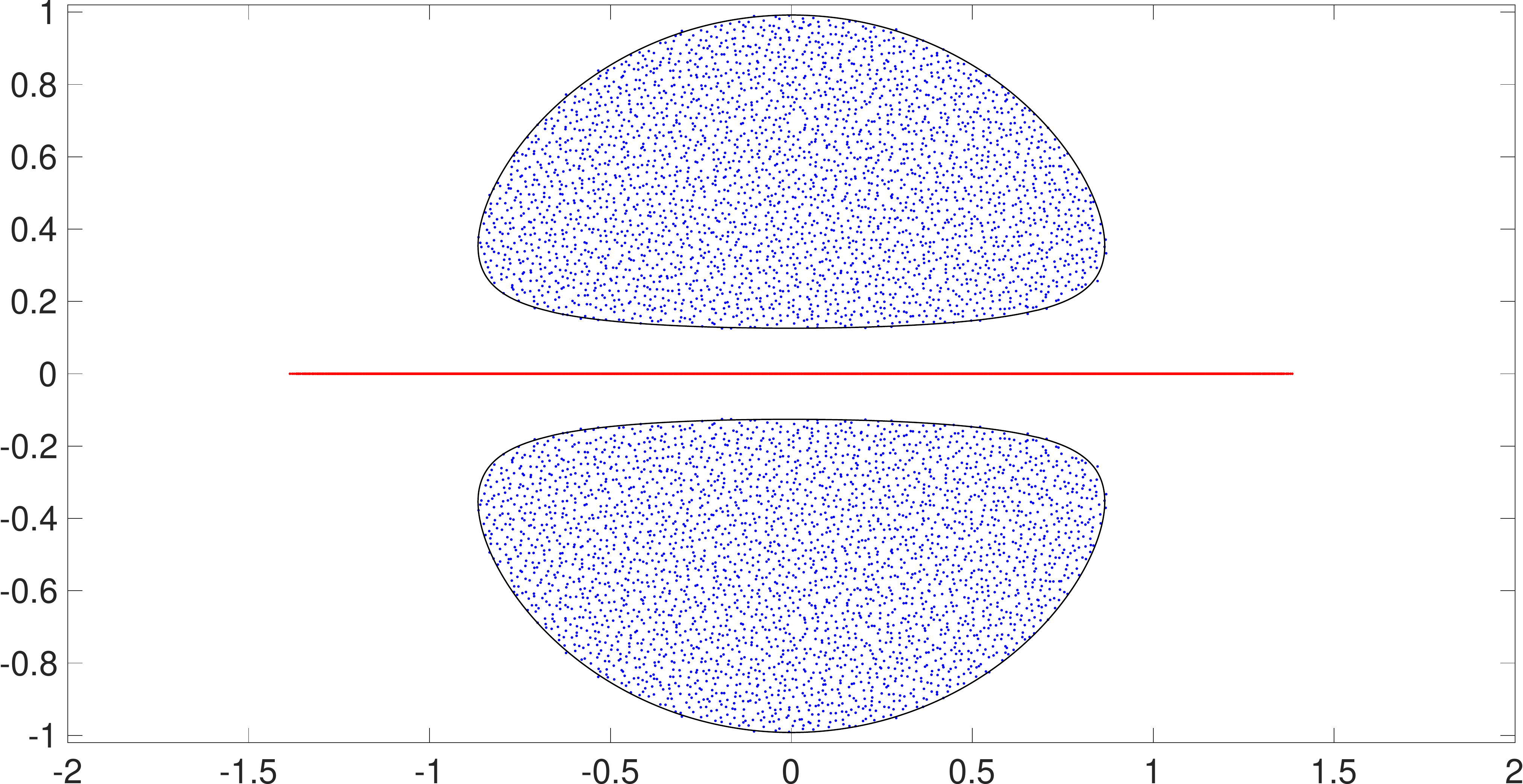}\hfill\includegraphics[width=0.495\textwidth]{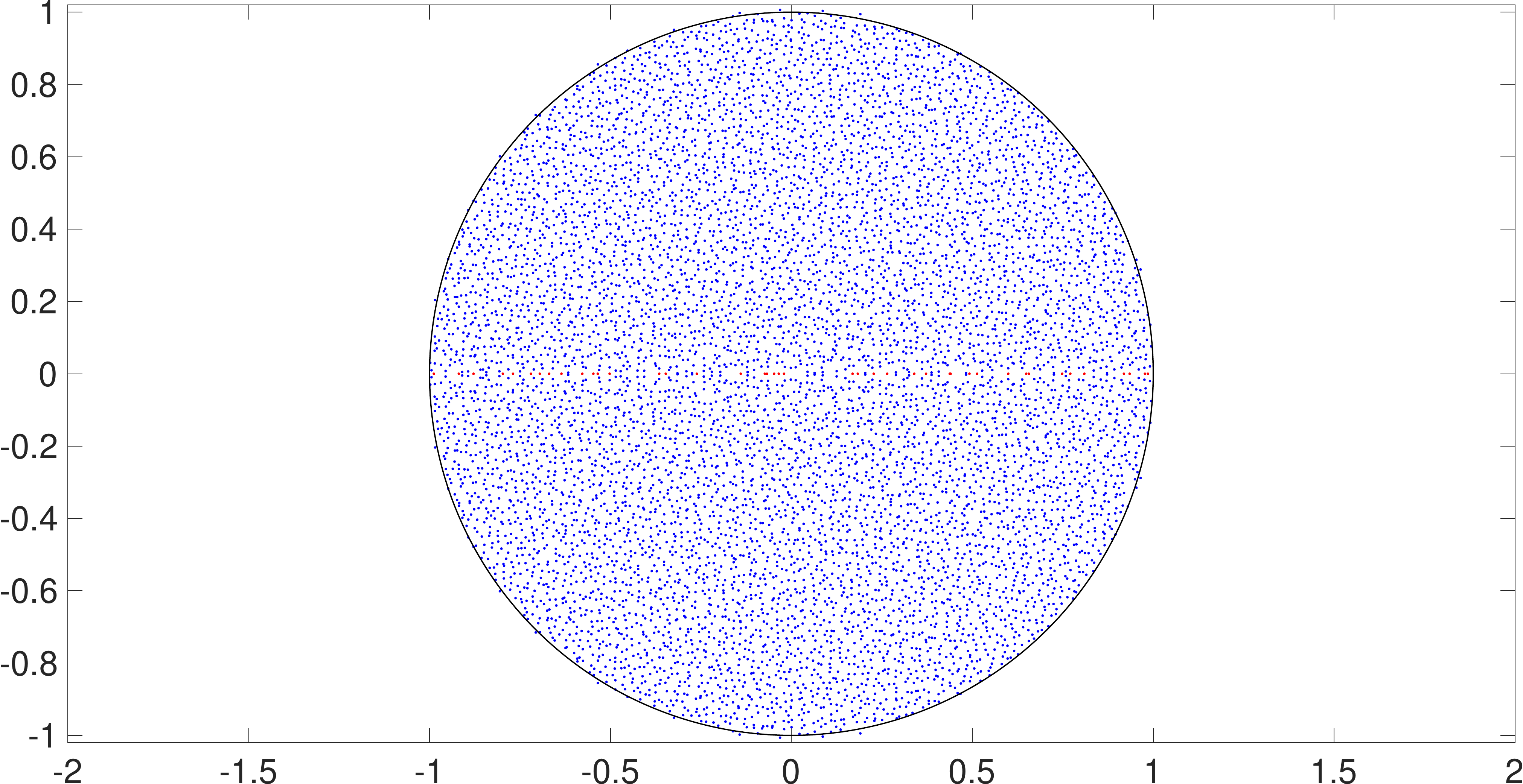} 
	\end{center} 
	\caption{Scatter plots showing the spectrum of $\PHI$ from one sample. Complex eigenvalues are marked in blue, real ones in red. $N=8192$ with $m=1$, $\lambda=1/8,1/4,3/8,1/2$ (upper left to lower right) }\label{fig:nonholomorphic}
\end{figure}
The scatter plots of complex eigenvalues in Figs.~\ref{fig:nonholomorphicN128} and \ref{fig:nonholomorphic} illustrate numerically the results for various values of $\lambda$. The blue dots correspond to complex eigenvalues of $\PHI$ and the red dots correspond to real eigenvalues. In Fig.~\ref{fig:nonholomorphicN128} we used 64 samples of matrix size $N=128$. Fig.~\ref{fig:nonholomorphic}, on the other hand,  was obtained from a single sample of matrix size $N=8192$. That a single large sample is enough to represent reliably the averaged density is the result of the self-averaging phenomenon, typical of the eigenvalue density of large random matrices. It demonstrates very well how uniformly the eigenvalues are distributed in the domain ${\mathcal D}$. The solid black lines in both figures represent the theoretical boundaries given in Eq.~\eqref{eq:rpm}. Finite-$N$ corrections are pronounced in Fig.~\ref{fig:nonholomorphicN128} as one can clearly see complex eigenvalues outside of the theoretical domain $\mathcal{D}$. In contrast, for $N=8192$ in Fig.~\ref{fig:nonholomorphic}, one can hardly see any outliers. Scatter plots for $N=32786$ can be found in \cite{FR-review}.  

On the lower right plot in Fig.~\ref{fig:nonholomorphicN128} and \ref{fig:nonholomorphic}, we see the degenerate case $\lambda=1/2$, for which the theoretical large-$N$ prediction \eqref{eq:rhoreal} (see also \eqref{eq:vshaped}) gives null density of real eigenvalues. Finite-$N$ corrections for $\lambda=1/2$ will result in small amounts of real eigenvalues, as is demonstrated very well by the corresponding plates in Figs.~\ref{fig:nonholomorphicN128} and \ref{fig:nonholomorphic}, and in accordance with Fig.~\ref{fig:vshaped}. In all other plots in Figs.~\ref{fig:nonholomorphicN128} and \ref{fig:nonholomorphic} we see pronounced distributions of real eigenvalues, with the red dots appearing to form a continuous line. This is so because they are very dense as there are $|N-2k|=N|1-2\lambda|$ real eigenvalues packed in the finite segment $[-x_0,x_0]$. For fixed $\lambda$, the typical distance between eigenvalues on the real lines is of order $N^{-1}$ while the mean distance between nearest neighbors of complex eigenvalues in the bulk is of order $N^{-1/2}$.   

For each sample, the eigenvalue distribution is symmetric with respect to the real axis. This is due to the fact that the characteristic polynomial of $\det(z-\PHI)$ has real coefficients and the complex eigenvalues appear in conjugate pairs. It is a manifestation of (broken) PT symmetry. The eigenvalue distribution has an additional reflection symmetry with respect to the imaginary axis, but only after averaging over all samples, because due to \eqref{eq:GUE} (or \eqref{ensemble}), $\PHI$ and $-\PHI$ are equi-probable.

As was mentioned above (following \eqref{eq:fracnonhol}), at $N=8192$ we can already see in Fig.~\ref{fig:nonholomorphic} that the complex eigenvalues are spread quite uniformly across the blobs, which agrees with our finding in \eqref{eq:rhoC} that the theoretical density there is constant. In Fig.~4 of \cite{FR-review} one can see two dimensional histograms which show that the density is very flat in the bulk. Only at the edges of those histograms one can notice some deviation from the constant density which is a typical finite $N$ size effect, that appears also in the Ginibre ensemble or in normal matrix models \cite{AKM,LR}.

From \eqref{eq:rpm} we find that the distance between the complex domain and the real axis (on both sides) is
\begin{equation}
r_-(\pi/2) = m^{-1} \sin(\theta_0/2).
\end{equation}
Thus, the two blobs kiss and touch the real axis when $\lambda$ approaches $1/2$. The latter is the degenerate case which we have mentioned above, and in which the complex eigenvalues fill uniformly a disk, like in the (real) Ginibre ensemble. This is of course the case $\tr\M{B}=0$ alluded to in \eqref{zerobh}, with $b\equiv0$  and $G(w) = 1/w$ throughout the holomorphic region  $r>1/m$.

\subsection{The Non-Compact Algebras $su(k,N-k)$ and $so(k,N-k)$, and Pseudo- and Anti-Pseudo-Hermiticity}
The metric \eqref{eq:metricB} is of special importance, because it appears in the definition of the non-compact Lie groups $U(k,N-k)$ and $O(k,N-k)$. By definition, the group $U(k,N-k)$ preserves the inner product $\M{u}^\dagger \M{B} \M{v}$ of vectors $\M{u},\M{v}\in \CC^N$, that is, 
\begin{equation}\label{U}
\M{g}^\dagger \M{B} \M{g} = \M{B},\quad \forall \M{g}\in U(k,N-k). 
\end{equation}
Upon expressing 
\begin{equation}\label{UAlgebra}
\M{g} = e^{i\PHI}
\end{equation}
in terms of elements $\PHI$ of the Lie algebra $u(k,N-k)$, we can rewrite \eqref{U} as 
\begin{equation}\label{U1}
e^{-i\PHI^\dagger} \M{B} e^{i\PHI} = \M{B},
\end{equation}
which is equivalent to the intertwining relation \eqref{intertwining} $\PHI^\dagger\M{B} = \M{B}\PHI$. Thus, the PH random matrix ensemble studied in this section is in effect an ensemble of elements of $u(k,N-k)$ in the large-$N, k$ limit, with fixed finite $\lambda = k/N$. 

In explicit terms, one can easily check that the matrices $\PHI$ which solve this intertwining relation are the {\em direct sum} of block-diagonal hermitian matrices and block-off-diagonal anti-hermitian matrices. The hermitian ones have diagonal blocks of dimensions $k\times k$ and $(N-k)\times (N-k)$, which generate compact $U(k)\otimes U(N-k)$ rotations within the two invariant subspaces. Their eigenvalues are purely real, of course. The anti-hermitian matrices have rectangular blocks of dimensions $k\times (N-k)$ and $(N-k)\times k$, with imaginary eigenvalues. They generate non-compact transformations (complexified boosts), which mix the two sub-spaces. In addition, due to their ``chiral" off-diagonal rectangular structure, they also possess at least $|N-2k|$ ``kinematical" zero modes which are typical of rectangular matrices. In ``physical" terms, think of the subspace orthogonal to such a complexified boost, which remains invariant. In other words, the corresponding boost generator annihilates these vectors. The full matrices $\PHI$, which dominate the ensemble and whose spectra was discussed in Sections \ref{Bhol} and \ref{Bnh}, are the direct sum of the hermitian compact generators and anti-hermitian non-compact generators. Thus, in essence, the formalism presented in this paper solves this non-trivial addition problem of the direct sum of block-diagonal hermitian random matrices and block-off-diagonal antihermitian random matrices. 

The random variable  $t=\frac{1}{N}\tr\PHI = \frac{1}{N}\tr \M{AB}$, which is  a weighted sum of the Gaussian variables in \eqref{eq:GUE}, follows itself the Gaussian distribution 
\begin{equation}\label{gaussian-trace}
p_N(t) = {Nm\over\sqrt{2\pi u}} e^{-{N^2m^2\over 2u}t^2},\quad u = \frac{1}{N} \tr \M{B}^2.
\end{equation}
Therefore (assuming a finite limit for $u$), 
\begin{equation}\label{large-N}
\lim_{N\to\infty} p_N(t)  = \delta(t),
\end{equation}
and $t$ becomes deterministically zero in the large-$N$ limit. Thus, more precisely, in this limit, our PH random matrix model is effectively an ensemble of the traceless part of $u(k,N-k)$, and our results in this section describe the spectral properties of elements of the Lie algebra $su(k,N-k)$ in the large-$N,k$ limit. 
The cases $k=0$ and $k=N$, that is $\M{B} = \pm\mathbb{1},$ correspond to $\PHI$ being a GUE matrix. Indeed, GUE can be thought of as an ensemble of elements of randomly chosen generators of the algebra $su(N)$.

In a similar manner, by definition, the group $O(k,N-k)$ preserves the inner product $\M{u}^T \M{B} \M{v}$ of vectors $\M{u},\M{v}\in \RR^N$, that is, 
\begin{equation}\label{O}
\M{g}^T \M{B} \M{g} = \M{B},\quad \forall \M{g}\in O(k,N-k). 
\end{equation}
Let us focus on the subgroup $SO(k,N-k)$ of $O(k,N-k)$. Upon expressing 
\begin{equation}\label{OAlgebra}
\M{g} = e^{\PHI}
\end{equation}
in terms of a real matrix $\PHI$, an element of the Lie algebra $so(k,N-k)$, we can rewrite \eqref{O} as 
\begin{equation}\label{U2}
e^{\PHI^T} \M{B} e^{\PHI} = \M{B},
\end{equation}
which is equivalent to the intertwining relation  
\begin{equation}\label{APS}
\PHI^T\M{B} =- \M{B}\PHI,
\end{equation}
meaning that $\PHI$ is anti-pseudosymmetric (APS) with respect to $\M{B}$. 

Similarly to the $su(k,N-k)$ case discussed above, one can easily check that such APS matrices $\PHI$ are the direct sum of block-diagonal real antisymmetric matrices and block-off-diagonal real symmetric matrices. The antisymmetric ones have diagonal blocks of dimensions $k\times k$ and $(N-k)\times (N-k)$, which generate compact $SO(k)\otimes SO(N-k)$ rotations within the two invariant subspaces. Their eigenvalues are purely imaginary, of course.  The real symmetric matrices have rectangular blocks of dimensions $k\times (N-k)$ and $(N-k)\times k$, with real eigenvalues. They generate non-compact transformations (boosts), which mix the two sub-spaces. In addition, due to their ``chiral" off-diagonal rectangular structure, they also possess at least $|N-2k|$ ``kinematical" zero modes, because boosts do not affect directions perpendicular to them. The full matrices $\PHI$, which dominate the ensemble, are the direct sum the antisymmetric compact generators and real symmetric non-compact generators.

If $\PHI$ is APS, the pure imaginary matrix $i\PHI$ is PH and satisfies the intertwining relation \eqref{intertwining}. 

In \cite{FRuniversality} we shall bring strong numerical evidence in support of universality of our results for the spectral density of the ensemble of PH matrices discussed in this section. In particular, we shall demonstrate that replacing the GUE ensemble \eqref{eq:GUE} with a properly rescaled GOE ensemble, that is, taking the matrices $\M{A}$ to be random real symmetric matrices, does not affect the large-$N$ results for the eigenvalue density of the real matrices $\PHI=\M{A}\M{B}$. In this case, $\PHI^T\M{B} = \M{B}\PHI$ are pseudo-symmetric with respect to $\M{B}$. 

Arguing along the lines of universality of the average spectra of real and complex PH hermitian matrices $\PHI$ in the large-$N$ limit, we should expect similar universality in the spectra of anti-PH real and complex matrices as well. Any complex anti-PH matrix can be written as $i\PHI$, with $\PHI$ a complex PH matrix. Thus the large-$N$ spectrum of complex anti-PH matrices is given by our results for PH matrices, rotated by $90$ degrees in the complex plane. Spectral universality thus tells us that the average spectrum of randomly chosen generators of $so(k,N-k)$ in the large-$N,k$ limit should be just the spectrum obtained in this section, rotated by $90$ degrees in the complex plane.

\section{Discussion} 
In this paper we have introduced a family of pseudo-hermitian random matrices as a new concept in random matrix theory. The conceptual novelty in these models is that PH matrices are hermitian with respect to a given {\em indefinite} metric. Major applications of pseudo-hermiticity include (and is certainly not limited to) the studying $PT$-symmetric quantum mechanical systems in the phase of broken $PT$-symmetry, as well as their classical gain-loss balanced analogs. Thus, in the usual spirit of Random Matrix Theory \cite{Mehta}, randomness of the PH matrices discussed in the present paper can model complicated or chaotic systems, or systems which are disordered to begin with. We shall study such applications of PH random matrix theory in future publications. 

 In this paper we have applied the diagrammatic method to derive the self-consistent gap equations in the planar limit, from which we determined the phase structure of the model in the complex plane. In particular, we have derived analytical expressions for the average Green's function of the model in the holomorphic and non-holomorphic phases as functions of the indefinite metric, which can be used to determine the average density of eigenvalues of such matrices both on the real axis and in the complex plane. As a concrete example, we have applied our formalism to studying a family of PH random matrix models for the generators of the classical non-compact Lie algebra $su(k,N-k)$  in the large-$N,k$ limit, with finite $\lambda = k/N$. We have calculated explicitly their average density of eigenvalues on the real axis and in the complex plane. We have also carried meticulous numerical analysis of these matrices, part of which was presented here. The numerical results agree very well with our analytical predictions. More analytical and numerical details pertaining to this model will be given in a forthcoming paper \cite{FRuniversality}, which will also demonstrate universality of our results for the density of eigenvalues. 
 
A preview of the main idea and results of this paper was published in \cite{FR-review}, which also reviewed our results on quasi-hermitian random matrices  and their application to the vibrational spectra of mechanical systems with large connectivity \cite{FRDecember, MK}, as well as analytical and numerical results pertaining to yet another PH random matrix model with very rich and interesting phase structure \cite{FR-Tmodel}. 

The results presented here and in \cite{FRDecember, MK, FR-review} are just the tip of the iceberg. Quasi- and pseudo-hermitian random matrices are clearly a promising new direction in the vast ocean of Random Matrix Theory.

\newpage
\setcounter{equation}{0}
\setcounter{section}{0}
\renewcommand{\theequation}{A.\arabic{equation}}
\renewcommand{\thesection}{Appendix}
\section{Technical and Mathematical Details}\label{appendix}
\vskip 5mm
\setcounter{section}{0}
\renewcommand{\thesection}{A}

\subsection{Symmetries of the spectrum of $\M{H}$ in \eqref{H2N}}\label{symmetries}
One can deduce the symmetries of the spectrum of $\M{H}$ in \eqref{H2N}  in the complex plane and its relation to the spectrum of the PH matrix $\PHI$ by analyzing the eigenproblem 
\begin{equation}\label{eigen}
\M{H}\psi = z\psi,
\end{equation}
directly, or alternatively, by studying the properties of the resolvent \eqref{resolvent}. Here we shall opt for the latter. 
 
Recall that $\PHI = \M{AB}$ and $\PHI^\dagger = \M{BA}$ are isospectral by virtue of \eqref{intertwining}, implying equality of the traces of the two diagonal blocks in \eqref{resolvent}, which we repeat here for convenience: 
\begin{equation}\label{resolvent-appendix}
\frac{1}{z-\M{H}}=\left(\begin{array}{cc} 
\frac{z}{z^2-\M{A}\M{B}}& \M{A}\frac{1}{z^2-\M{B}\M{A}}\\
\M{B}\frac{1}{z^2-\M{A}\M{B}}& \frac{z}{z^2-\M{B}\M{A}}
\end{array}\right).
\end{equation}
Thus, 
\begin{equation}\label{traces}
\tr_{(2N)} {1\over z-\M{H}}  = \tr_{(N)} {2z\over z^2-\PHI}  = \sum_{i=1}^N \left({1\over z-\sqrt{w_i}} + {1\over z+\sqrt{w_i}}\right)\,.
\end{equation}
We see that  each eigenvalue $w_i$ of $\PHI$ leads to a pair of eigenvalues $\pm\sqrt{w_i}$ of $\M{H}$.  Moreover, as was mentioned in the Introduction, isospectrality of $\PHI$ and $\PHI^\dagger$ also means that their complex eigenvalues come in complex conjugate pairs. Thus, if $\Im w_i\neq 0$, then $w_i^*$ is also an eigenvalue of $\PHI$ (distinct from $w_i$), and therefore $\pm\sqrt{w_i^*}$ are eigenvalues of $\M{H}$ as well. Therefore, while the PH nature of $\PHI$ renders its spectrum symmetric under reflection with respect to the real axis, the spectrum of $\M{H}$ has four-fold reflection symmetry, with respect to both the real and imaginary axes.

Another way to deduce this symmetry, without resorting directly to the spectral properties of $\PHI$, is to depart from the chiral block structure of \eqref{H2N} which implies 
\begin{equation}\label{anticommutaiton}
\{\M{H},\M{\Gamma\}} = 0\,,
\end{equation}
where 
\begin{equation}\label{chiral}
\M{\Gamma} = \M{1}_N\otimes\sigma_3 = \left(\begin{array}{cc} 
\M{1}_N& 0\\
0& -\M{1}_N
\end{array}\right),
\end{equation} 
is the chirality matrix.  Thus, if $\psi$ is a (right) eigenvector of $\M{H}$ with eigenvalue $z$, then $\M{\Gamma}\psi$ is also an eigenvector of $\M{H}$, with eigenvalue $-z$. This accounts for the reflection symmetry $z\leftrightarrow -z$ of the spectrum of $\M{H}$ through the origin. Moreover, as can be seen from \eqref{resolvent-appendix}, the LHS of \eqref{traces} is invariant under interchanging $\M{A}\leftrightarrow\M{B}$, which from \eqref{H2N} is equivalent to interchanging $\M{H}\leftrightarrow\M{H}^\dagger$. Thus, the spectrum of $\M{H}$ is invariant under complex conjugation as well, and therefore under reflection with respect to the real axis. 

In the large-$N$ limit, the eigenvalues  of $\M{H}$ typically become dense in a two-dimensional domain $\widetilde{\mathcal D}$ in the complex plane. More precisely, as it turns out, due to the PH nature of $\PHI$, the spectral domain of $\M{H}$ also contains a one-dimensional density component of eigenvalues concentrated in finite segments  along the real and imaginary axes, the union of which we denote by $\tilde\sigma$. We have shown here that the domain $\widetilde{\mathcal D}\cup \tilde\sigma$ must have four-fold reflection symmetry with respect to both the real and imaginary axes.

\subsection{The Diagonal Blocks of the Resolvent \eqref{eq:G1}}\label{sec:diagonals}
In this section we will establish a certain functional relation between the diagonal blocks of the resolvent \eqref{eq:G1}, which hold true for every realization of the random matrix $\M{A}$, that is, before taking averages. To this end we evaluate \eqref{eq:G1} at $\eta = is$, with $s\in\RR$, that is, 
\begin{equation}\label{G1s} 
\hat{\mathcal{G}}(is; z, z^*)= 
\left(\begin{array}{cc} {-is \over s^2 + (z-\M{H})(z-\M{H})^\dagger} & (z-\M{H}) {1\over s^2 + (z-\M{H})^\dagger (z-\M{H}) }\\ (z - \M{H})^\dagger {1\over s^2 + (z-\M{H})(z-\M{H})^\dagger} & {-is \over s^2 + (z-\M{H})^\dagger (z-\M{H})}\end{array}\right).
\end{equation}
Let us now substitute $\M{H}$ from \eqref{H2N} in \eqref{G1s} and start by computing the hermitian and positive definite inverse
\begin{eqnarray}\label{inverse1} 
{1\over s^2 + (z-\M{H})(z-\M{H})^\dagger} &=& \left(\begin{array}{cc} s^2 + |z|^2 +\M{A}^2 & -z\M{B} - z^*\M{A}\\ -z^*\M{B}-z\M{A} & s^2 + |z|^2 +\M{B}^2\end{array}\right)^{-1}\nonumber\\{}\nonumber\\&\equiv& \left(\begin{array}{cc} P(s;z,z^*) & Q(s;z,z^*)\\ Q^\dagger(s;z,z^*) & R(s;z,z^*)\end{array}\right)
\end{eqnarray}
and find the corresponding $N\times N$ blocks as 
\begin{eqnarray}\label{PQR}
P(s;z,z^*) &=& P^\dagger(s;z,z^*)  = \left[s^2 + |z|^2 + \M{A}^2 - (z\M{B} + z^*\M{A}){1\over s^2 + |z|^2 + \M{B}^2}(z^*\M{B}+z\M{A} )\right]^{-1}\nonumber\\{}\nonumber\\
R(s;z,z^*) &=& R^\dagger(s;z,z^*)  = \left[s^2 + |z|^2 + \M{B}^2 - (z\M{A} + z^*\M{B}){1\over s^2 + |z|^2 + \M{A}^2}(z^*\M{A}+z\M{B} )\right]^{-1}\nonumber\\{}\nonumber\\
Q(s;z,z^*) &=& \nonumber\\ &&  \hspace{-1.5cm}{1\over s^2 + |z|^2 + \M{A}^2 - (z\M{B} + z^*\M{A}){1\over s^2 + |z|^2 + \M{B}^2}(z^*\M{B}+z\M{A} )}(z\M{B} + z^*\M{A}){1\over s^2 + |z|^2 + \M{B}^2}\,.
\end{eqnarray}
As should be clear from the structure of the second matrix in \eqref{inverse1}, by interchanging $\M{A}\leftrightarrow\M{B}$ we have $P(s;z,z^*)\leftrightarrow R(s;z,z^*)$ and $Q(s;z,z^*)\leftrightarrow Q^\dagger(s;z,z^*)$. (To see that interchanging  $\M{A}$ and $\M{B}$ in the last expression in \eqref{PQR} has the effect of transforming $Q$ to its adjoint requires a little bit of reshuffling of the various matrix factors.) 

Similarly, we find
\begin{eqnarray}\label{inverse2} 
{1\over s^2 + (z-\M{H})^\dagger(z-\M{H})} &=& \left(\begin{array}{cc} s^2 + |z|^2 +\M{B}^2 & -z\M{B} - z^*\M{A}\\ -z^*\M{B}-z\M{A} & s^2 + |z|^2 +\M{A}^2\end{array}\right)^{-1}\nonumber\\{}\nonumber\\&=& \left(\begin{array}{cc} R(s;z^*,z) & Q^\dagger(s;z^*,z)\\ Q(s;z^*,z) & P(s;z^*,z)\end{array}\right)\,,
\end{eqnarray}
that is, the result \eqref{inverse1} after simultaneously interchanging $z\leftrightarrow z^*$ and $\M{A}\leftrightarrow\M{B}$.

By collecting our results from equations \eqref{inverse1}-\eqref{inverse2}, substituting them in \eqref{G1s} and using the definition of the $N\times N$ blocks \eqref{blocks}, we thus conclude that 
\begin{eqnarray}\label{diagonals} 
\left(\begin{array}{cc} \hat{\mathcal{G}}_{11}(is;z,z^*)  & \hat{\mathcal{G}}_{12}(is;z,z^*)  \\
\hat{\mathcal{G}}_{21}(is;z,z^*)  & \hat{\mathcal{G}}_{22}(is;z,z^*) \end{array}\right) &=&  -is \left(\begin{array}{cc} P(s;z,z^*) & Q(s;z,z^*)\\ Q^\dagger(s;z,z^*) & R(s;z,z^*)\end{array}\right)
\nonumber\\{}\nonumber\\
\left(\begin{array}{cc} \hat{\mathcal{G}}_{33}(is;z,z^*)  & \hat{\mathcal{G}}_{34}(is;z,z^*)  \\
\hat{\mathcal{G}}_{43}(is;z,z^*)  & \hat{\mathcal{G}}_{44}(is;z,z^*) \end{array}\right)&=& - is  \left(\begin{array}{cc} R(s;z^*,z) & Q^\dagger(s;z^*,z)\\ Q(s;z^*,z) & P(s;z^*,z)\end{array}\right)\,.
\end{eqnarray}
Upon tracing each $N\times N$ block in these equations, we also obtain the corresponding expressions for the unaveraged block traces 
\begin{equation}\label{blocktrace-unaverage}
\alpha\beta (is;z,z^*)  = {1\over N}\tr \hat {\cal G}_{\alpha\beta}(is;z,z^*),\quad \alpha,\beta = 1,2,3,4
\end{equation}
(whose averaged versions appear in \eqref{blocktrace}) in terms of traces over $P$, $Q$ and $R$.

In particular, the diagonal blocks $ \hat{\mathcal{G}}_{11},  \hat{\mathcal{G}}_{22},  \hat{\mathcal{G}}_{33}$ and $ \hat{\mathcal{G}}_{44}$ are all anti-hermitian $N\times N$ matrices, and are related by 
\begin{equation}\label{interrelation1}
\hat{\mathcal{G}}_{44}(is;z,z^*)  = \hat{\mathcal{G}}_{11}(is;z^*,z)
\end{equation}
and 
\begin{equation}\label{interrelation2} 
\hat{\mathcal{G}}_{33}(is;z,z^*) =  \hat{\mathcal{G}}_{22}(is;z^*,z), 
\end{equation}
and correspondingly 
\begin{equation}\label{trace-interrelation1}
{44}(is;z,z^*)  ={11}(is;z^*,z)
\end{equation}
and 
\begin{equation}\label{trace-interrelation2} 
{33}(is;z,z^*) = {22}(is;z^*,z)
\end{equation}
for their {\em purely imaginary} traces.

The two diagonal blocks of \eqref{G1s} are unitarily equivalent\footnote{This can be seen from their singular value decompositions. Equality of the traces of these two blocks can be also established by expanding each of these blocks in inverse powers of $s^2$ and using cyclicity of the trace.} and therefore have equal traces.  As can be seen from \eqref{G1s}, in the limit $s\rightarrow 0$ we have
\begin{eqnarray}\label{diagonals0} 
\lim_{s\to 0\pm}\left(\begin{array}{cc} \hat{\mathcal{G}}_{11}(is;z,z^*)  & \hat{\mathcal{G}}_{12}(is;z,z^*)  \\
\hat{\mathcal{G}}_{21}(is;z,z^*)  & \hat{\mathcal{G}}_{22}(is;z,z^*) \end{array}\right) &=& \mp i\pi \delta\left(\sqrt{(z-\M{H})(z-\M{H})^\dagger}\right)  
\nonumber\\{}\nonumber\\
\lim_{s\to 0\pm}\left(\begin{array}{cc} \hat{\mathcal{G}}_{33}(is;z,z^*)  & \hat{\mathcal{G}}_{34}(is;z,z^*)  \\
\hat{\mathcal{G}}_{43}(is;z,z^*)  & \hat{\mathcal{G}}_{44}(is;z,z^*) \end{array}\right)&=&  \mp i\pi\delta\left(\sqrt{(z-\M{H})^\dagger (z-\M{H})}\right),
\end{eqnarray}
(the averaged versions of which appear in \eqref{etas}). 
We thus obtain 
\begin{eqnarray}\label{trace-deltas} 
11(i0\pm;z,z^*)  + 22(i0\pm;z,z^*)  &=& \mp i{\pi\over N} \tr \delta\left(\sqrt{(z-\M{H})(z-\M{H})^\dagger}\right)  
\nonumber\\{}\nonumber\\
33(i0\pm;z,z^*)  + 44(i0\pm;z,z^*) &=&  \mp i{\pi\over N} \tr \delta\left(\sqrt{(z-\M{H})^\dagger (z-\M{H})}\right).
\end{eqnarray}
As was mentioned above, these two traces are equal, so that 
\begin{equation}\label{equal-sums}
11(i0\pm;z,z^*)  + 22(i0\pm;z,z^*)  = 33(i0\pm;z,z^*)  + 44(i0\pm;z,z^*) \,.
\end{equation}
Clearly, the sums of diagonal block traces on the LHS of \eqref{trace-deltas} do not vanish if and only if $z$ is an eigenvalue of $\M{H}$ (and therefore $z^*$ an eigenvalue of $\M{H}^\dagger$). Thus, they serve as sort of order parameters indicating the location of the support $\widetilde{\mathcal D}\subset\CC$ of the density of eigenvalues of $\M{H}$ in the large-$N$ limit \cite{FZ}. 

The discussion in (section 3 of) \cite{FZ} focused on non-hermitian matrices $\M{H}$ with purely two-dimensional spectral domain ${\widetilde{\mathcal D}}$. For example, if $\M{H}$ is drawn from Ginibre's ensemble \cite{ginibre}, normalized in the large-$N$ limit to have the unit disk as its spectral domain, then 
\begin{equation}\label{ginibre}
\left\langle i{\pi\over N}\tr \delta\left(\sqrt{(z-\M{H})^\dagger (z-\M{H})}\right) \right\rangle =\left\{\begin{array}{c}  i \sqrt{1-|z|^2},\quad |z|<1\\ 0,\quad\quad\quad\quad\quad |z|>1\end{array}\right.,
\end{equation}
which should hold true to high precision even before averaging over the ensemble, due to the self-averaging property of large random matrices. Going beyond Ginibre's ensemble to general rotationally invariant ensembles of non-hermitian matrices, the Single Ring Theorem \cite{FZ1} asserts that the spectral domain $\widetilde{\mathcal D}$ of $\M{H}$ is either a disk or an annulus, and for such models as well, one can compute the diagonal trace on the LHS of \eqref{ginibre} (see Section 4 in \cite{FZ1}) and show that it vanishes only outside $\widetilde{\mathcal D}$. Thus, this diagonal trace indeed fulfills its role as order parameter indicating the location of the two-dimensional spectral domain. 

However, as was mentioned at the end of Section \ref{symmetries},  the spectral domain of $\M{H}$ in \eqref{H2N} also contains a one-dimensional eigenvalue density component $\tilde\sigma$, located symmetrically along segments of the real and imaginary axes. We now show that this one-dimensional spectral component has a negligible effect on the order parameter, producing a benign removable discontinuity, invisible in the large-$N$ limit. To get oriented, take as a start $\M{H}$ to be a {\em hermitian} matrix - the opposite extreme of a Ginibre-type matrix - whose eigenvalues $x_1,\ldots , x_N$ condense in the large-$N$ limit in some finite segment $\tilde\sigma$ along the real axis, with density $\rho^{(1)}(x)$ as in \eqref{rho1}. In this case, it is straightforward to show that 
\begin{eqnarray}\label{hermitian1}
&& \left\langle i{\pi\over N}\tr \delta\left(\sqrt{(z-\M{H})^\dagger (z-\M{H})}\right) \right\rangle\nn\\
&&\qquad =i\pi\int\limits_{\tilde\sigma} \rho^{(1)}(x')\delta\left(\sqrt{(x-x')^2+y^2}\right) \dd x'= \left\{\begin{array}{ll}  i\pi \rho^{(1)}(x),& y= 0\\ 0,& y\neq 0.\end{array}\right.
\end{eqnarray}
As a function of $y$, this diagonal trace is null, except a removable discontinuity at $y=0$, and fails to serve as an order parameter indicating the position of $\tilde\sigma$.  This conclusion holds true also in the more generic case, when the spectral domain of $\M{H}$ is a union of both a one-dimensional part $\tilde\sigma$ and a two-dimensional part $\widetilde{\mathcal D}$. 

In order to proceed, let us carry a singular value decomposition of the matrix $z-\M{H}$ 
\begin{equation}\label{SVD}
z-\M{H} = \M{U}\M{\Lambda}\M{V}
\end{equation}
where $\M{U}(z)$ and $\M{V}(z)$ are unitary $2N\times 2N$ matrices and 
\begin{equation}\label{Lambda}
\M{\Lambda} = {\rm diag}(\Lambda_1(z), \ldots, \Lambda_{2N}(z))
\end{equation}
are the corresponding singular values (all positive, by definition). We shall assume a generic matrix $\M{H}$, with non-degenerate eigenvalue spectrum $z_1, \ldots z_{2N}$. Let's pick some ordering of these eigenvalues. By appropriate choice of the unitary matrices, we can always order the singular values to follow the ordering of the corresponding eigenvalues. Clearly, ${\rm rank}(z-\M{H}) = {\rm rank}(\M{\Lambda}(z))$, and therefore when the parameter $z$ hits one of these non-degenerate eigenvalues, only the corresponding singular value would vanish
\begin{equation}\label{vanishing1}
\lim_{z\to z_k} \Lambda_k(z) = 0.
\end{equation}
In this case, we immediately see that $\M{V}^\dagger(z_k)\hat {\bf e}_k$ is an eigenvector on the right of $\M{H}$, belonging to the eigenvalue $z_k$, and $\M{U}(z_k)\hat {\bf e}_k$ is the corresponding eigenvector on the left (with $\hat {\bf e}_k$ the $k$th Cartesian unit vector).

By substituting \eqref{SVD} in \eqref{trace-deltas}, we see that the two equal diagonal traces on the RHS of \eqref{trace-deltas}
\begin{equation}\label{SVDos}
{1\over N} \tr \delta\left(\sqrt{(z-\M{H})^\dagger(z-\M{H})}\right)  =  {1\over N} \tr \delta\left(\sqrt{(z-\M{H})(z-\M{H})^\dagger}\right)  = {1\over N} \sum_{k=1}^{2N} \delta \left(\Lambda_k(z)\right)
\end{equation} 
are in fact the density of {\em singular values} of $z-\M{H}$ evaluated at zero. Thus, our diagonal traces count the density of null singular values of $z-\M{H}$ given $z$, and in light of \eqref{vanishing1}, contain information about the density of complex eigenvalues at that point in the complex plane. To make progress, we need to determine how $\Lambda_k(z)$ vanishes as $z\to z_k$. This we can infer by looking at 
\begin{equation}\label{detH}
\det(z-\M{H})  = \prod_{k=1}^{2N}(z-z_k),
\end{equation}
and therefore
\begin{equation}\label{detLambda}
\det\left((z-\M{H})^\dagger (z-\M{H})\right)  = \prod_{k=1}^{2N} \Lambda^2_k(z) = \prod_{k=1}^{2N}|z-z_k|^2.
\end{equation}
Thus, by tuning $z = z_k + \epsilon$, where $|\epsilon|\ll$ the smallest gap between the eigenvalues of $\M{H}$, we immediately conclude that 
\begin{equation}\label{Lambdak}
\Lambda_k(z_k + \epsilon)  = c_k |z-z_k| (1+{\mathcal O}(\epsilon))
\end{equation}
where the positive coefficient is 
\begin{equation}\label{ck}
c_k = \prod_{\substack{l=1\\ l\neq k}}^{2N} {|z_k-z_l| \over \Lambda_l(z_k)}.
\end{equation}
By substituting this result in \eqref{SVDos}, we thus conclude that 
\begin{eqnarray}\label{SVDos1}
&&\hspace{-2cm}{1\over N}\tr \delta\left(\sqrt{(z-\M{H})^\dagger(z-\M{H})}\right)  =  {1\over N}\sum_{k=1}^{2N} {1\over c_k} \delta (|z-z_k|) = {1\over N}\!\!\!\!\sum_{\substack{\rm complex\\ \rm eigenvalues}} {1\over c_k} \delta (|z-z_k|) + \nonumber\\{}\nonumber\\
&+&\!\!\!\!\!\!\!\!\!\!  {1\over N}\!\!\!\!\!\!\sum_{\substack{\rm real\\ \rm eigenvalues}} {1\over c_k} \delta \left(\sqrt{(x-x_k)^2 + y^2}\right) + {1\over N}\sum_{\substack{\rm imaginary\\ \rm eigenvalues}} {1\over c_k} \delta \left(\sqrt{x^2 + (y-y_k)^2}\right) 
\end{eqnarray} 
When we set, for example, $y\neq 0$ and go off the real axis, purely real eigenvalues make no contribution to the trace, while for $y=0$, in the large-$N$ limit, they only give rise to a removable discontinuity, because then their contribution converges to a smooth function, as in \eqref{hermitian1}. Obviously, a similar conclusion holds for the purely imaginary component in $\tilde\sigma$. 
Finally, note that for $\M{H}$ a normal matrix, and in particular, a hermitian matrix, clearly $\Lambda_k(z) = |z-z_k|$, and therefore $c_k=1 \,\forall k$, which is consistent in the large-$N$ limit with \eqref{hermitian1}.

Thus, to summarize the discussion of the few previous paragraphs, the sums of diagonal block traces on the LHS of \eqref{trace-deltas} serve as order parameters indicating the location of the {\em two-dimensional} support $\widetilde{\mathcal D}\subset\CC$ of the density of eigenvalues of $\M{H}$ in the large-$N$ limit, and are effectively insensitive to the one-dimensional part $\tilde\sigma$ of the spectral domain. 

A corollary of this assertion is that both these sums {\em  must} vanish (modulo the removable discontinuity along $\tilde\sigma$) outside $\widetilde{\mathcal D}$, that is, in the complementary domain $\widetilde{\mathcal D}^c = \CC/\widetilde{\mathcal D}$. All the ${\alpha\alpha}(is=0; z,z^*)$'s in \eqref{trace-deltas} are of the same sign (depending on whether $s\rightarrow 0+$ or $0-$). Therefore, vanishing of such a sum implies vanishing of each of its terms separately. 
We conclude that all four ${\alpha\alpha}(is=0; z,z^*)$ must vanish in $\widetilde{\mathcal D}^c$.

The complementary domain $\widetilde{\mathcal D}^c$ obviously shares the same four-fold reflection symmetry with $\widetilde{\mathcal D}$, discussed in the previous Appendix \ref{symmetries}. Therefore, if some $z\in \widetilde{\mathcal D}^c$, then also $z^*\in \widetilde{\mathcal D}^c$, leading to the conclusions that 
\begin{equation}\label{vanishing}
{\alpha\alpha}(is=0; z,z^*) =0 \Leftrightarrow {\alpha\alpha}(is=0; z^*,z) =0\,, \alpha=1,2,3,4 \quad \forall z\in \widetilde{\mathcal D}^c\,,
\end{equation}
which is consistent with \eqref{trace-interrelation1} and \eqref{trace-interrelation2}.

In contrast, at this point, before taking averages, we cannot determine whether each of the four diagonal traces ${\alpha\alpha}(i0;z,z^*)\neq 0$ separately for $z\in \widetilde{\mathcal D}$. All we can say at this point is that the two equal sums in \eqref{equal-sums} do not vanish $\forall z\in \widetilde{\mathcal D}$. (The identities \eqref{trace-interrelation1} and \eqref{trace-interrelation2} we were able to establish, merely interchange $z\leftrightarrow z^*$ in \eqref{equal-sums}.) We will be able to answer this question in the affirmative (see \eqref{AOC} in the main text) only after averaging over $\M{A}$.

\subsection{Derivation of the Self Energy in the Planar Limit}\label{SE-details}
Let us start by deriving a useful elementary algebraic identity. To this end we introduce the standard basis vectors in matrix space 
\begin{equation}\label{basis}
(\hat{\M{e}}_{ij})_{kl} = \delta_{ik}\delta_{jl}, 
\end{equation}
namely, the $N\times N$ matrix whose all entries are null, except the $ij$-th element. Then, it is straightforward to show that (repeated indices are summed over)
\begin{equation}\label{matrix}
\left(\hat{\M{e}}_{ij}\M{M}\hat{\M{e}}_{ji}\right)_{pq} = \delta_{pq}\,{\rm tr}\M{M},
\end{equation}
for any $N\times N$ matrix $\M{M}$.

In order to proceed, we write the inverse of \eqref{eq:G} as  
\begin{equation}\label{inverseG}
\hat{\mathcal{G}}^{-1} = \hat{\mathcal{G}}_0^{-1} - \hat{\mathcal{A}}\,, 
\end{equation} 
where $\hat{\mathcal{G}}_0^{-1}$ was defined in \eqref{bareprop} and where the fluctuating Gaussian part is 
\begin{equation}\label{fluctA} 
\hat{\mathcal{A}} = \left(\begin{array}{cccc}
0&0&0& \M{A}\\
0&0&0&0\\
0&0&0&0\\
 \M{A}&0&0&0\end{array}\right)  = \M{A}\otimes (\sigma_+\otimes\sigma_+ + \sigma_-\otimes\sigma_-).
\end{equation} 
In the last equation 
\begin{equation}\label{sigmapm}
\sigma_+ = \left(\begin{array}{cc}
0&1\\
0&0\\
\end{array}\right)\quad{\rm and}\quad  \sigma_- = \left(\begin{array}{cc}
0&0\\
1&0\\
\end{array}\right)
\end{equation}
are the usual raising and lowering Pauli matrices.  
 
According to Dyson and Schwinger, the self-energy, associated with our Gaussian fluctuating matrix \eqref{fluctA}, can be expressed in the large-$N$ planar limit in terms of the full propagator as 
\begin{equation}\label{SE-def}
\hat\Sigma = \langle \hat{\mathcal{A}} \langle\hat{\mathcal{G}}\rangle\hat{\mathcal{A}} \rangle_c
\end{equation}
where the subscript $c$ indicates the connected second order cumulant \eqref{cumulant}. Thus,  
\begin{eqnarray}\label{SE-calc}
\hat\Sigma &=&\langle A_{ij}\hat{\M{e}}_{ij}\otimes (\sigma_+\otimes\sigma_+ + \sigma_-\otimes\sigma_-)  \langle\hat{\mathcal{G}}\rangle A_{kl}\hat{\M{e}}_{kl}\otimes (\sigma_+\otimes\sigma_+ + \sigma_-\otimes\sigma_-)\rangle_c\nonumber\\
&=&\langle A_{ij} A_{kl}\rangle_c \, \hat{\M{e}}_{ij}\otimes (\sigma_+\otimes\sigma_+ + \sigma_-\otimes\sigma_-)  \langle\hat{\mathcal{G}}\rangle \hat{\M{e}}_{kl}\otimes (\sigma_+\otimes\sigma_+ + \sigma_-\otimes\sigma_-)\nonumber\\
&=&{1\over Nm^2} \delta_{il}\delta_{jk} \, \hat{\M{e}}_{ij}\otimes (\sigma_+\otimes\sigma_+ + \sigma_-\otimes\sigma_-)  \langle\hat{\mathcal{G}}\rangle \hat{\M{e}}_{kl}\otimes (\sigma_+\otimes\sigma_+ + \sigma_-\otimes\sigma_-)\nonumber\\
&=& {1\over Nm^2}  \, \hat{\M{e}}_{ij}\otimes (\sigma_+\otimes\sigma_+ + \sigma_-\otimes\sigma_-)  \langle\hat{\mathcal{G}}\rangle \hat{\M{e}}_{ji}\otimes (\sigma_+\otimes\sigma_+ + \sigma_-\otimes\sigma_-).
\end{eqnarray}

Multiplying through, and displaying the blocks \eqref{blocks} of $\langle\hat{\mathcal{G}}\rangle $ on the way, we thus obtain 

\begin{eqnarray}\label{SE-final}
\hat\Sigma &=& {1\over Nm^2} \left(\begin{array}{cccc}
\hat{\M{e}}_{ij}\langle\hat{\mathcal{G}}_{44}\rangle\hat{\M{e}}_{ji} &0&0&\hat{\M{e}}_{ij}\langle\hat{\mathcal{G}}_{41}\rangle\hat{\M{e}}_{ji} \\
0&0&0&0\\
0&0&0&0\\
\hat{\M{e}}_{ij}\langle\hat{\mathcal{G}}_{14}\rangle\hat{\M{e}}_{ji} &0&0&\hat{\M{e}}_{ij}\langle\hat{\mathcal{G}}_{11}\rangle\hat{\M{e}}_{ji} 
\end{array}\right)\nonumber\\{}\nonumber\\
&=&{1\over m^2} \M{1}_N\otimes \left(\begin{array}{cccc}
\overline{44} &0&0&\overline{41} \\
0&0&0&0\\
0&0&0&0\\
\overline{14} &0&0&\overline{11}
\end{array}\right)\,,
\end{eqnarray}
where we have used the definition \eqref{blocktrace}.

\subsection{Detailed Derivation of the Gap Equation at $\eta=0$}\label{gap-details}
We start from Eqs. \eqref{propagator} and \eqref{SE} and set $\eta=is = 0$. We can then rewrite these equations as 
\begin{equation}\label{inverseprop}
\langle \hat{\mathcal{G}}\rangle_{|_{s=0}}^{-1}=\left(\hat{\mathcal{G}}_0^{-1}-\hat{\Sigma}\right)_{|_{s=0}} = \left(\begin{array}{cccc} 
-{\overline{44}\over m^2}&0&z&-{\overline{41}\over m^2}\\
0&0&-\M{B}&z\\
z^*&-\M{B}&0&0\\
-{\overline{14}\over m^2}&z^*&0&-{\overline{11}\over m^2}
\end{array}\right)_{|_{s=0}} 
\end{equation} 
where all quantities are calculated at $\eta=is = 0$ and where we have suppressed explicit indication of the $N\times N$ unit matrix $\M{1}_N$ in the appropriate blocks.  

It would be more convenient at this point to rearrange the blocks of the matrix on the right hand side of \eqref{inverseprop} such that all average block traces appear in one diagonal block. To this end we act with the unitary permutation matrix 
\begin{equation}\label{permutaiton}
S =  \left(\begin{array}{cccc} 
1&0&0&0\\
0&0&1&0\\
0&0&0&1\\
0&1&0&0
\end{array}\right)
\end{equation}
on the blocks of \eqref{inverseprop} and obtain 
\begin{eqnarray}\label{inverseprop1}
S^{-1}\langle \hat{\mathcal{G}}\rangle_{|_{s=0}}^{-1} S = S^{-1}\left(\hat{\mathcal{G}}_0^{-1}-\hat{\Sigma}\right)_{|_{s=0}} S &=& \left(\begin{array}{cccc} 
-{\overline{44}\over m^2}&-{\overline{41}\over m^2}&0&z\\
-{\overline{14}\over m^2}&-{\overline{11}\over m^2}&z^*&0\\
0&z&0&-\M{B}\\
z^*&0&-\M{B}&0
\end{array}\right)_{|_{s=0}}\nonumber\\{}\nonumber\\
&=& \left(\begin{array}{cccc} 
a&b&0&z\\
b^*&c&z^*&0\\
0&z&0&-\M{B}\\
z^*&0&-\M{B}&0
\end{array}\right)
\end{eqnarray} 
where we have used \eqref{1144} and the notations of \eqref{abc}.
Let us now invert both sides of \eqref{inverseprop1} and write 
\begin{equation}\label{prop1}
S^{-1}\langle \hat{\mathcal{G}}\rangle_{|_{s=0}} S = \left(\begin{array}{cccc} 
a&b&0&z\\
b^*&c&z^*&0\\
0&z&0&-\M{B}\\
z^*&0&-\M{B}&0
\end{array}\right)^{-1}
\equiv\left(\begin{array}{cc} 
X&Y\\
W&Z\end{array}\right).
\end{equation} 
After a straightforward but tedious calculation, we find the $2N\times 2N$ blocks on the RHS of \eqref{prop1} as\footnote{Here and in subsequent places, all $N\times N$ blocks commute with each other. Therefore we are free to factor them in and out of the block matrices $X,Y,W$ and $Z$ as if they were scalars.}
\begin{eqnarray}\label{XYWZ}
X &=& {1\over ac - |b+z^2 \M{B}^{-1}|^2} \left(\begin{array}{cc} 
c& - (b+z^2 \M{B}^{-1})\\
-(b^*+z^{*2} \M{B}^{-1})&a\end{array}\right)\nonumber\\{}\nonumber\\
Y &=& {\M{B}^{-1}\over ac - |b+z^2 \M{B}^{-1}|^2} \left(\begin{array}{cc} 
cz& -z^* (b+z^2 \M{B}^{-1})\\
-z(b^*+z^{*2} \M{B}^{-1})&az^*\end{array}\right)\nonumber\\{}\nonumber\\
W &=&  {\M{B}^{-1}\over ac - |b+z^2 \M{B}^{-1}|^2} \left(\begin{array}{cc} 
cz^* & - z^* (b+z^2 \M{B}^{-1})\\
-z(b^*+z^{*2} \M{B}^{-1})&az\end{array}\right)\nonumber\\{}\nonumber\\
Z &=&  {\M{B}^{-1}\over ac - |b+z^2 \M{B}^{-1}|^2}\left(\begin{array}{cc} 
c|z|^2\M{B}^{-1}& -ac + b^* (b+z^2 \M{B}^{-1})\\
-ac +b(b^*+z^{*2} \M{B}^{-1})&a|z|^2\M{B}^{-1}\end{array}\right),\nonumber\\
\end{eqnarray}
where we have used the abbreviated notation introduced in \eqref{abbreviation}. Note that the product $ac$ is real and negative, being the product of two pure imaginary quantities of equal signs.

The permutation matrix $S$ reshuffles the blocks of $ \hat{\mathcal{G}}$ on the left-hand side of \eqref{prop1} into 
\begin{equation}\label{prop2}
S^{-1}\langle \hat{\mathcal{G}}\rangle_{|_{s=0}} S = \left\langle\left(\begin{array}{cccc}
\hat{\mathcal{G}}_{11}&\hat{\mathcal{G}}_{14}&\hat{\mathcal{G}}_{12}&\hat{\mathcal{G}}_{13}\\
\hat{\mathcal{G}}_{41}&\hat{\mathcal{G}}_{44}&\hat{\mathcal{G}}_{42}&\hat{\mathcal{G}}_{43}\\
\hat{\mathcal{G}}_{21}&\hat{\mathcal{G}}_{24}&\hat{\mathcal{G}}_{22}&\hat{\mathcal{G}}_{23}\\
\hat{\mathcal{G}}_{31}&\hat{\mathcal{G}}_{34}&\hat{\mathcal{G}}_{32}&\hat{\mathcal{G}}_{33}\\
\end{array}\right)\right\rangle 
\end{equation}
By comparing this equation with the first equation in \eqref{limitsOD}, we see that after reshuffling with $S$, all the four blocks which comprise $\langle{1\over z-\M{H}}\rangle$ migrate to the lower left  corners of the blocks $X, Y, W$ and $Z$. Note also that the four diagonal blocks remain on the main diagonal after this reshuffling.

The desired gap equation is obtained by equating \eqref{prop1} and \eqref{prop2}. In particular, by equating the upper $2N\times 2N$ diagonal block of \eqref{prop2} with the corresponding block $X$ of \eqref{prop1}, given explicitly by the first equation in \eqref{XYWZ}, we obtain
\begin{equation}\label{14gap}
\left\langle\left(\begin{array}{cc}
\hat{\mathcal{G}}_{11}&\hat{\mathcal{G}}_{14}\\
\hat{\mathcal{G}}_{41}&\hat{\mathcal{G}}_{44}\\\end{array}\right)\right\rangle  =  {1\over ac - |b+z^2 \M{B}^{-1}|^2} \left(\begin{array}{cc} 
c& - (b+z^2 \M{B}^{-1})\\
-(b^*+z^{*2} \M{B}^{-1})&a\end{array}\right).
\end{equation}
Thus, by equating block traces on both sides of this equation, and with the definition \eqref{abc} in mind, we obtain a set of self-consistent equations for the quantities $a, b$ and $c$:
\begin{eqnarray}\label{abc-selfconsistent-appendix}
a &=& -{a\over Nm^2} \tr  {1\over ac - |b+z^2 \M{B}^{-1}|^2} \nonumber\\{}\nonumber\\
c &=& -{c\over Nm^2} \tr  {1\over ac - |b+z^2 \M{B}^{-1}|^2} \nonumber\\{}\nonumber\\
b &=& {1\over Nm^2} \tr  {b^* + z^{*2}\M{B}^{-1}\over ac - |b+z^2 \M{B}^{-1}|^2}.  
\end{eqnarray}
Block tracing the remaining block in \eqref{14gap} just produces the complex conjugated last equation in \eqref{abc-selfconsistent-appendix}, as it should.  

Two of the diagonal blocks of $\langle\hat{\mathcal {G}}\rangle$ appear in \eqref{14gap}. Let us record here explicitly also the two remaining diagonal blocks
\begin{eqnarray}\label{12gap}
\langle\hat{\mathcal{G}}_{22}\rangle &=&  {c|z|^2 \M{B}^{-2}\over ac - |b+z^2 \M{B}^{-1}|^2} \nonumber\\
\langle\hat{\mathcal{G}}_{33}\rangle &=&    {a|z|^2 \M{B}^{-2}\over ac - |b+z^2 \M{B}^{-1}|^2}  
\end{eqnarray}
which we read off \eqref{prop1}-\eqref{prop2}.

After solving for $a, b$ and $c$ from \eqref{abc-selfconsistent-appendix} and feeding these quantities back to \eqref{XYWZ}, we can  determine all the $N\times N$ blocks $\langle\hat{\mathcal{G}}_{\alpha\beta}\rangle$ of $\langle\hat{\mathcal{G}}\rangle$. As we can see from \eqref{limitsOD}, of particular interest among all these blocks is 
\begin{equation}\label{31-appendix}
\langle\hat{\mathcal{G}}_{31}\rangle = \left\langle\frac{z}{z^2-\M{A}\M{B}}\right\rangle, 
\end{equation}
which is essentially the desired resolvent of $\PHI = \M{A}\M{B}$. Thus,  from \eqref{XYWZ} and \eqref{prop2} we obtain
\begin{equation}\label{31-final-appendix}
\langle\hat{\mathcal{G}}_{31}\rangle = - {z\M{B}^{-1}(b^*+z^{*2} \M{B}^{-1})\over ac - |b+z^2 \M{B}^{-1}|^2}.
 \end{equation}

\vspace{1cm}
\section*{Acknowledgements}

This research was supported by the Israel Science Foundation (ISF) under grant No. 2040/17.  Computations presented in this work were performed on the Hive computer cluster at the University of Haifa, which is partly funded by ISF grant 2155/15. Finally, we thank T.~Can for turning our attention to \cite{carlson}.

\end{document}